\documentclass[a4paper,usenames,dvipsnames,11pt]{article}
\usepackage{cite}
\usepackage{tabularx,booktabs,multirow}
\usepackage[makeroom]{cancel}
\usepackage{amssymb}
\usepackage{amsmath}
\usepackage{booktabs, cellspace, hhline}
\usepackage{multirow}
\usepackage{makecell}
\usepackage{graphicx}

\usepackage{hyperref}
\usepackage{xcolor}
\usepackage{setspace}
\usepackage{colortbl}
\newcommand{\LC}{\textrm{LC}}
\newcommand{\NLC}{\textrm{NLC}}
\newcommand{\FC}{\textrm{FC}}

\usepackage{authblk}
\usepackage{mleftright}

\usepackage[mathscr]{euscript}

\newcommand{\NC}{N_{\text{\tiny C}}}

\newcommand{\M}{\mathcal{M}}
\newcommand{\A}{\mathcal{A}}
\newcommand{\dd}{\text{d}}

\newcommand{\gen}{gen2$\to$3}

\begin{document}

\title{\textbf{Leading-colour-based unweighted event generation for multi-parton tree-level processes}}

\date{}
\author{
Rikkert Frederix$^{1\,}$\footnote{E-mail:  \texttt{rikkert.frederix@fysik.lu.se}},
Timea Vitos$^{2,3}$\footnote{E-mail:  \texttt{timea.vitos@physics.uu.se}},\\
{\small\it $^{1}$ Department of Physics, Lund University,} 
\\%
{\small\it S\"olvegatan 14A, SE-223 62, Lund, Sweden}\\
{\small\it $^{2}$ Department of Physics and Astronomy, Uppsala University,} 
\\%
{\small\it Box  516,  751 20, Uppsala, Sweden  }\\
{\small\it $^{3}$ Institute for Theoretical Physics, ELTE  E\"otv\"os Lor\'and  University } 
\\%
{\small\it P\'azm\'any  P\'eter  s\'et\'any  1/A,  H-1117  Budapest,  Hungary}\\
}
\maketitle

\begin{abstract}
    \noindent In this work, we revisit unweighted event generation
  for multi-parton tree-level processes in massless QCD. We introduce
  a two-step approach, in which initially unweighted events are generated at
  leading-colour (LC) accuracy, followed by a reweighting of these events
  to full-colour (FC) accuracy and applying an additional
  unweighting cycle. This method leverages the simple structure of LC
  integrands, enabling optimized phase-space parameterisations and
  resulting in high primary unweighting efficiencies, ranging from the
  percent level for $2 \to 4$ processes to the per-mille level for $2
  \to 7$ processes. Given that the LC-accurate matrix elements closely approximate
  the FC-accurate ones, the secondary unweighting efficiencies exceed 50\%. Our
  results suggest that this two-step approach offers an efficient
  alternative to direct event generation at FC accuracy.

\end{abstract}
\thispagestyle{empty}
\vfill

\newpage

\begingroup
\hypersetup{linkcolor=black}
\tableofcontents
\endgroup

\section{Introduction}
In preparation for the large amount of data to be collected during the
upcoming high-luminosity phase of the Large Hadron Collider (LHC),
improvements and developments in accuracy and precision of the theoretical description of both
background and signal processes are essential. Because of the nature of hadronic
collisions and the intrinsic properties of quantum chromodynamics (QCD),
interesting (signal) events are often accompanied by multiple well-separated hadronic
jets---either as part of the signal, or as a background to it. Predictions for
cross sections and differential distributions for such processes are
time-consuming to compute, even at leading order (LO) in perturbation theory,
since they entail high-dimensional ($3n-2$, with $n$ the number of final-state
particles) phase-space integrals of integrands whose complexity increases
factorially with $n$. Ideally, these predictions come in the form of
\emph{unweighted events}, meaning events distributed according to a probability density function proportional to the magnitude of the integrand. In
practice, the only viable options to perform these integrals are numerical
(Monte Carlo) methods, combined with a hit-or-miss step to obtain unweighted
events.

In order to improve the efficiency of the integration, techniques such as
importance
sampling~\cite{Lepage:1977sw,Lepage:2020tgj,vanHameren:2007pt}, and
multi-channeling~\cite{Kleiss:1994qy,Ohl:1998jn} are
employed~\cite{Maltoni:2002qb,Papadopoulos:2000tt,Krauss:2001iv,Kilian:2007gr,Gleisberg:2008fv,Alwall:2014hca,Sherpa:2019gpd}. Moreover,
determining the optimal parametrisation of the phase space with respect to the integration variables is crucial for efficient integration. Generating the momenta
uniformly~\cite{Kleiss:1985gy,Platzer:2013esa} is typically not desirable;
parametrisations in which integration variables are linked to invariants that
enter the propagator structure of the Feynman diagrams contributing to the
process under consideration~\cite{Byckling:1969luw,Byckling:1969sx} yield greater integration
efficiency. Alternatively to using the
propagator structure, approaches presented in
Refs.~\cite{Draggiotis:2000gm,vanHameren:2002tc} use the antenna structure of
the QCD matrix elements as part of the phase-space parametrisation. More
recently, parametrisations that align the most-common variables on which
physical phase-space cuts are applied with integration variables have been
proposed~\cite{Bothmann:2023siu}. Going beyond the analytic transformations to
flatten the integrand, multivariate machine learning
algorithms are also under development to improve the efficiency in phase-space
integration~\cite{Bendavid:2017zhk,Klimek:2018mza,Chen:2020nfb,Gao:2020vdv,Bothmann:2020ywa,Gao:2020zvv,Danziger:2021eeg,Heimel:2022wyj,Janssen:2023ahv,Heimel:2023ngj,Deutschmann:2024lml,Heimel:2024wph}.

In addition to improving the integration, to address the factorial growth in complexity of
the integrand, techniques involving recursion relations and/or Monte Carlo
sampling of helicities and colours of the external particles have been
employed~\cite{Berends:1987me,Caravaglios:1995cd,Caravaglios:1998yr,Draggiotis:1998gr,Mangano:2002ea,Duhr:2006iq,Gleisberg:2008fv,Mattelaer:2021xdr}.

In this work, we revisit the method of unweighted event generation for
multi-parton tree-level processes in QCD. We propose an approach in which an
approximation of the complete LO-accurate matrix elements is used in a first, integration,
step. Unweighted events are generated in this approximation. As a second step,
these generated events are reweighted to the exact LO-accurate matrix elements, and a
secondary unweighting step is performed.

Our paper is structured as follows. In Sec.~\ref{sec:method} we discuss the
details of our method, i.e., the approximation used for the integrand, as well
as the four different phase-space parametrisations that we have considered for the integration step of our method. In Sec.~\ref{sec:results} we present our findings for the
primary and secondary unweighting efficiencies in an idealised, fixed-energy partonic
collision as well as in a more realistic LHC setup, taking parton distribution functions into account. In the final
section we conclude and give a short outlook on the results presented.

\section{Two-step event generation}\label{sec:method}
Multi-parton matrix elements are slow to evaluate, they have non-trivial
structures, and the corresponding phase space has a high dimensionality, resulting in a large
variance in the weights of the generated phase-space points. For the
generation of unweighted events, i.e., events that are distributed according
to the density of the integrand (and hence all assigned the same weight, or,
contribution, to the integrand evaluation), the weight of each event needs to
be compared to the maximum possible weight and is kept with a probability
equal to the ratio of its weight to the maximum weight. The fraction of
events kept in this way is what is commonly referred to as the unweighting
efficiency. Since there can easily be multiple orders of magnitude between
typical event weights and the maximum weight, the unweighting efficiencies for
multi-parton processes can be quite low, which is problematic when the
evaluation time of the matrix elements is long.

The idea behind our method is to split this event generation procedure into
two (or more) steps. In the first step, an approximation of the integrand
is used that reduces it to a simple form, making it very fast to evaluate,
and unweighted events are generated from this integrand. Besides the
requirement that the evaluation time of the approximated integrand must be short, using
a simple integrand has the additional benefit that it is easier to find a good
phase-space parametrisation that reduces the variance in the generated
weights, resulting in an improved unweighting efficiency. In the subsequent
step(s), these low-accuracy events are reweighted to higher
accuracy\footnote{Our work is focused on tree-level accuracy within
  perturbative QCD, and thus a full accuracy is never obtained. Here, high-
  and full-accuracy refer to the level of approximation of the integrand,
  which is, by itself, a tree-level approximation of the scattering
  process.}. For these reweight factors, the full (tree-level)
time-consuming matrix elements need to be evaluated. However, only for the
smaller set of phase-space points corresponding to the unweighted events
generated in the first step. The resulting weighted events can be unweighted
again by keeping them with a probability equal to the ratio of their reweight
factors to the maximum possible reweight factor, ensuring that the final set of
events with unit weights follows the exact distribution of the initial integrand.

For this method to work efficiently, the approximation must fulfill
some basic properties. Firstly, it needs to be fast to evaluate,
and/or an efficient parametrisation of the phase space must exist such
that peaks in the integrand can be flattened accordingly. Furthermore,
the reweight factors which are needed to go from the approximation to the full-accuracy
events must have a small variance, i.e., the approximation cannot be
too inaccurate and must closely follow the full integrand in all of phase space. If
the variance is too large, the secondary unweighting of the
full-accuracy events can be rather inefficient, resulting in a poor
overall performance. It would be particularly problematic if the approximation grossly underestimated the full-accurate matrix elements in a particular
phase-space region, resulting in extremely large reweight factors,
which would completely spoil the efficiency of the secondary
unweighting. Specifically, if the approximation results in vanishing values in certain regions of phase space, the reweight factor would diverge in those cases. However, the approximation does not have to get
the overall normalisation of the full-accuracy result right: a
constant offset does not affect the secondary unweighting efficiency.

The general process we consider is the scattering of two incoming partons into
$n$ final-state partons which are labeled according to
\begin{equation}
a b \rightarrow 1 2 \ldots n.
\end{equation}
The integral which is to be computed for this process is the integral in the
phase space over the squared matrix elements, averaged over initial colour and
helicity and summed over all external assignments, convoluted with the two
parton distribution functions (for a hadronic collision setup), and including the flux factor, schematically,
\begin{equation}
\sigma_{a b \rightarrow 1 2 \ldots n} = \int |\M|^2 \dd\Phi,
\end{equation}
where we have suppressed the flux factor and the parton luminosity
factors. The phase-space measure includes the Bjorken-$x$ variables,
$\dd\Phi= \dd x_a \dd x_b\dd\Phi_n$, with $\dd\Phi_n$ denoting the usual
Lorentz-invariant phase-space measure for $n$ final-state partons.

\subsection{The approximation of the matrix elements}
As the approximation of the integrand under consideration, we use the
leading-colour (LC) approximation of the matrix elements, obtained by performing a large-$\NC$ ($\NC$ being the number of colours) expansion. Since the structure of the matrix elements at this accuracy is rather simple, efficient phase-space generators can be developed in this approximation. As shown later, the LC matrix elements form an excellent
approximation of the integral at full-colour (FC) accuracy, and result in a narrow weight
distribution for the reweight factors.

\subsubsection{Leading-colour matrix elements}\label{sec.LC}
The amplitudes appearing in the integrand are summed over colour and
helicity configurations of the external particles. They can be decomposed into
colour-ordered amplitudes, which are dependent on the colour ordering of the
external particles.  In general, the colour ordering can be expressed as an
ordered string of the particle labels, such as $(a123b45)$ for $n=5$ final-state particles. In this ordering,
if quarks are present in the process, a quark label always appears first in
the order and an anti-quark label always last. Depending on the particle
content, this sequence can be cyclically invariant (all-gluon processes),
fixed (for one-quark line processes), or redundant in the exchange of a
substring that begins with a quark label and ends with an anti-quark label
(for processes with multiple quark lines).  At leading colour, all interferences
between different colour-ordered amplitudes vanish. Schematically, using,
e.g.~the decomposition of the matrix elements based on the fundamental
basis\footnote{There are other bases, such as the multiplet
  basis~\cite{Keppeler:2012ih}, that are also diagonal at full
  colour. However, for those cases, the basis elements are no longer single
  colour-ordered amplitudes, and therefore much slower to evaluate.},
\begin{align}\label{eq.LC}
\int |\M|^2\dd\Phi = \int \sum_{i,j} \A_i C_{ij}\A^*_j \dd\Phi \quad
\xrightarrow{\LC} \quad & \int \sum_i \A_i C_{ii}\A^*_i
\dd\Phi = \sum_{i} C_{ii} \int |\A_i|^2 \dd\Phi,
\end{align}
where, in the final step, we have taken the sum outside of the
integral sign, which emphasizes that each term in the sum can be
integrated separately.
Note that for processes with (at least) two different-flavour quark lines, all
the elements on the diagonal of the colour matrix are considered in our LC
approximation as the notation in Eq.~\eqref{eq.LC} suggests, even though only
half of them contribute at LC and the other half starts only at
next-to-leading-colour (NLC) accuracy in the colour expansion,
see~e.g.~Ref.~\cite{Frederix:2021wdv}. In general, the number of elements in
the sum grows factorially with the number of external
particles. However, many elements in the sum yield the same result
since the phase space is symmetric for identical particles, and, in
practice, the number of separate integrals one needs to consider
scales, at worst, polynomially with the number of external
particles~\cite{Frederix:2021wdv,Badger:2012pg}. Moreover, using
recursion relations, such as the ones introduced in
e.g.~Refs.~\cite{Berends:1987me,Britto:2004ap}, a single colour ordering
for a given helicity configuration can be computed with, at most,
exponential scaling with the number of external
particles.

For the parton multiplicities we consider (up to $n \simeq 7$), it is
more efficient to explicitly sum over all helicities of the
colour-ordered amplitudes for each phase-space point, instead of
sampling over them. The reason is that sampling over helicities
increases the variance of the matrix elements in the phase-space
integration, leading to a poorer performance and decreased unweighting
efficiencies\footnote{This has been directly checked and verified.}.  Before writing the unweighted LC-accurate events to the event file, we assign them one helicity configuration randomly, weighted by
their relative contributions. The subsequent reweighting steps are
therefore performed for a single helicity configuration.

\subsubsection{Phase-space generation}\label{sec.PS}
\begin{figure}[htb!]
  \begin{center}
    \includegraphics[scale=0.5]{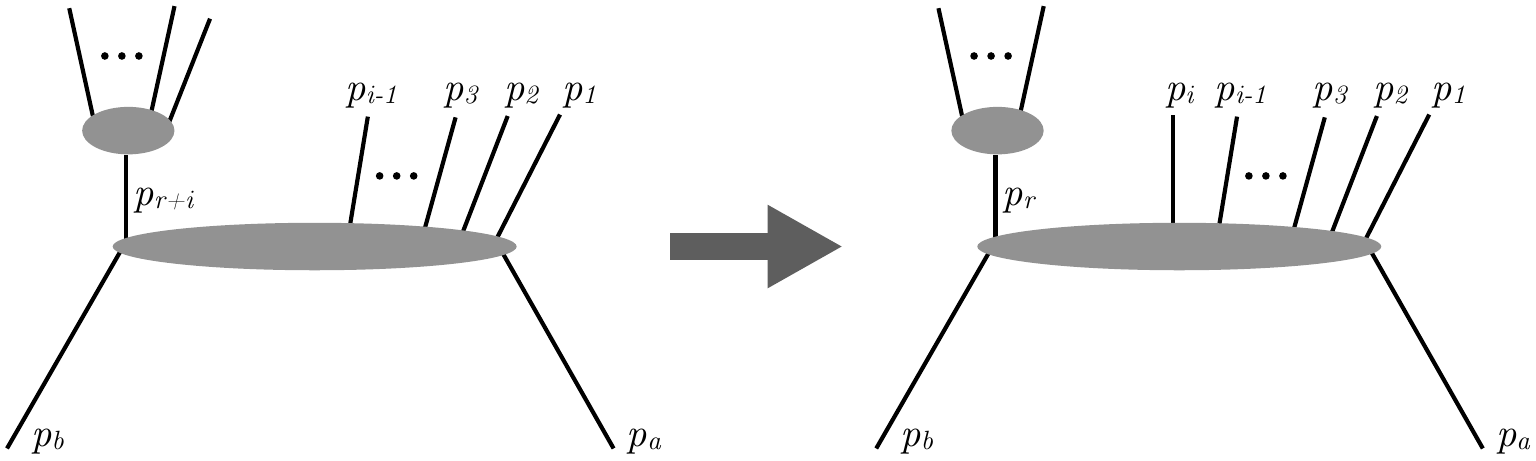}
    \caption{The step to generate the momentum of particle $i$ in the
      sequential phase-space generation: a pseudo-particle with
      momentum $p_{r+i}$ is split into a (pseudo-)particle with
      momentum $p_r$ and a particle with momentum $p_i$.}
    \label{PhaseSpaceSplit}
  \end{center}
\end{figure}

The general $n$-body phase-space measure,
including the two initial-state Bjorken-$x$ momentum fraction factors
relevant for hadron-hadron collisions, is proportional to
\begin{equation}
\dd x_a \dd x_b \dd \Phi_n = \dd x_a \dd x_b \Bigg(\prod_{i=1}^n\dd^4
p_i\delta(p^2_i-m_i^2)\Theta(E_i)\Bigg)\delta^{(4)}\Big(p_a+p_b-\sum_{i=1}^np_i\Big),
\end{equation}
where $p_a$ and $p_b$ are the four-momenta of the right- and left-incoming particles, and $p_i$ and $m_i$ are the four-momenta and masses
of the $n$ final-state particles. For simplicity, in the expression we have suppressed factors of
$2\pi$. For an implementation in a numerical code, the delta-functions
need to be resolved. Typically, but not necessarily, this is done by
combining each of the $\delta(p^2_i-m_i^2)$ factors with the
corresponding $\dd^4 p_i$ measures into $\dd^4
p_i\delta(p^2_i-m_i^2)=\frac{\dd^3\mathbf{p}_i}{2 E_i}$, and the four-dimensional
delta-function imposing overall momentum conservation,
$\delta^{(4)}\Big(p_a+p_b-\sum_{i=1}^np_i\Big)$, is used to express
the momentum of, without loss of generality, particle $n$ in terms of the momenta of
the other particles (as well as one of the
integration variables relevant for particle $n-1$). Most-conveniently, this allows for a sequential
generation of the phase space, whereby, typically, first the momenta
of the two incoming particles are generated by throwing random numbers
for $x_a$ and $x_b$, and then using $(n-2)$ universal blocks, each
consisting of three random variables to generate the momenta of
particles $1\ldots n-2$ by splitting, for each particle $i$, the
remaining momentum $p_{r+i} \to p_r + p_i$, as depicted in Fig.~\ref{PhaseSpaceSplit}. The last particle $n-1$ can be generated using a
slightly modified building block, containing only $2$ random
variables, while the momentum of particle $n$ follows directly from
momentum conservation once all the momenta of particles $1\ldots n-1$
are known.

In the universal blocks the three random variables can, in principle,
directly correspond to the Cartesian $x$, $y$, and $z$ components of the
momentum $p_i$, but it is more efficient to perform a change of variables and
relate these to variables that are directly relevant in matrix elements. In
particular one would like the peaks in the integrand of the phase-space
integral to align with the integration variables, such that a (non-uniform)
generation of the random variables (either through one-dimensional analytic
transformations or numerical methods such as the Vegas
algorithm~\cite{Lepage:1977sw,Lepage:2020tgj}) can flatten the peaks,
which reduces the variance among the weights of the generated
phase-space points and improves unweighting efficiency.

For the processes that we consider in this work, the strongest peaks follow
the antenna structure of the colour-ordered QCD amplitudes\footnote{While the
  actual divergence needs to be removed by phase-space cuts, dominant
  contributions to the integral still come from the phase-space regions
  close to these cuts.}. This is particularly evident in the maximally
helicity violating (MHV) amplitudes for the all-gluon process, that take a
particularly simple form~\cite{Parke:1986gb}, e.g.,
\begin{equation}\label{eq:mhv}
|\A_i|^2 \propto \frac{(p_a\cdot p_b)^4}{(p_a\cdot p_b) (p_b\cdot
  p_1)(p_1\cdot p_2)\cdots (p_n\cdot p_a)}\,,
\end{equation}
corresponding to particles $a$ and $b$ having opposite helicity to particles
$1,\ldots, n$ and corresponding to the colour ordering $(ab1\ldots n)$.  For
non-MHV helicity configurations, the amplitudes take on slightly more
complicated forms, see, e.g.~Ref.~\cite{Dixon:2010ik}, but the overall
structure remains. Therefore, the natural ordering in which to generate the
phase-space splittings, c.f.~Fig.~\ref{PhaseSpaceSplit}, is according to the
colour ordering\footnote{Also for processes with quarks, we take the (or, if
  there are several, one of the) naive
  colour ordering(s) (and assume invariance under cyclic permutations) as the sequence
  in the phase-space generation. This is reasonable since colour-ordered
  amplitude with quarks have fewer peaks than all-gluon
  amplitudes.}.  If the two incoming particles are adjacent in the colour
ordering under consideration, we generate the phase space in the same sequence
as the colour ordering. If, on the other hand, the two incoming particles are
not adjacent, we first perform an overall $p_a+p_b \to p_{r_1} +p_{r_2}$
phase-space generation, in which $p_{r_1}$ corresponds to the combined momenta
of the particles in the colour ordering between particles $a$ and $b$, and
$p_{r_1}$ to the combined momenta between particles $b$ and $a$. For each of
these sets, we then perform the splittings $p_{r_{1,2}}\to p_{r_{1,2}-i}+p_i$
sequentially, following the colour ordering as depicted in Fig.~\ref{PhaseSpaceSplit}.

We consider the following four phase-space parametrisations, each of
which uses different integration variables in the universal
blocks. 
\begin{enumerate}
\item The \textbf{Haag} phase-space generation
  method~\cite{vanHameren:2002tc} has been designed to generate the
  phase-space density directly according to the all-gluon MHV matrix
  elements, Eq.~\eqref{eq:mhv}. In the universal blocks used for the momentum split
  $p_{r+i} \to p_r+p_i$, this parametrisation uses rescaled invariants between the
  momenta $p_i$ and $p_{i-1}$ and the momenta $p_r$ and $p_a$ (or
  $p_b$), as well as the invariant mass of $p_r$, as the three
  integration variables (with the latter omitted to generate the
  momentum of particle $n-1$). In order to get the correct overall density, the
  generation of these variables must be performed following specific
  densities~\cite{vanHameren:2002tc}. 
  
\item The \textbf{t-channel} method~\cite{Byckling:1969sx} uses
  building blocks for the $p_{r+i} \to p_r+p_i$ momentum split in
  which the invariant computed from the momenta $p_r$ and $p_a$ (or
  $p_b$), i.e., the ``$t$-channel'' momentum, and the invariant mass of $p_r$
  are used as integration variables. The third variable is the
  azimuthal angle of $p_i$, defined in the frame where the momentum
  $p_{r+i}$ is at rest. Since, in the high-energy limit, the matrix
  elements are dominated by $t$-channel structures, see also
  e.g.~Ref.\cite{Andersen:2011hs}, it is expected that this
  parametrisation works well for multi-particle processes.
  
\item An extension to the t-channel method, to increase the number of
  invariants used as integration variables, was first described in
  Ref.~\cite{Byckling:1969luw}. This method is similar to the t-channel method
  in that it uses the invariant mass of the remaining momentum and the
  invariant computed from the current momentum with one of the
  initial-state momenta as integration variables. On the other hand,
  the azimuthal angle is replaced by the invariant mass between the
  current and the previously generated momentum $(p_i+p_{i-1})^2$. Since
  it requires there to be two final-state (pseudo-)particles, the ones
  with momentum $p_{r+i}$ and $p_{i-1}$ (see Fig.~\ref{PhaseSpaceSplit}),
  to generate the third momentum $p_i$,
  we have dubbed this method \textbf{\gen}.  It uses very similar
  variables as the Haag method, such that all the invariants appearing
  in the MHV antenna structure, Eq.~\eqref{eq:mhv}, can be mapped directly to integration
  variables\footnote{With a couple of exceptions: for $2 \to 2$ scattering at fixed partonic energy,
    there are only two independent integration variables, while the
    denominator in Eq.~\eqref{eq:mhv} contains four invariants (which are
    obviously not independent). For higher multiplicities, we generate the
    phase space separately for the final-state particles between the two
    incoming particles $a$ and $b$ and particles $b$ and $a$ in the colour
    ordering. If one of these two sets contains exactly one or two particles, the
    invariants related to that set are not all directly mapped to integration
    variables.}. However, it is not enforced that the density of the
  generated momenta exactly corresponds to the MHV antenna structure,
  rather, it relies on (numerical) importance sampling, as implemented
  in the Vegas algorithm, to generate momenta as close as possible
  to the density of the integrand.  This has the advantage that the
  algorithm is much simpler than the Haag one. Even though this method
  was developed in the late 1960s, this is, to the best of our
  knowledge, the first time that this phase-space parametrisation has
  been implemented and used to generate the phase space for general
  multi-particle processes.
  
\item The $\textbf{p}_T$\textbf{-based} method \cite{Mangano:2002ea,Bothmann:2023siu}, uses the transverse-momentum,
  the rapidity and the azimuthal angle of particle $i$ as
  integration variables in the universal blocks (all in the laboratory
  frame). Since each of the blocks are independent, it can be seen as
  a base-line method that does not incorporate any of the structure of the
  colour-ordered antennae in the phase-space generator. On the
  other hand, it has the advantage that the LHC analyses typically impose
  analysis cuts on transverse momenta and rapidities of final-state
  objects, which are aligned with the integration variables in this
  method. Therefore, importance-sampling grids can also optimise precisely
  close to these phase-space boundaries. Moreover, contrary to
  the previous three methods, this method does not start with
  generating the momenta of the two incoming particles. Rather, the
  integration variables relevant to the two Bjorken $x$'s are
  transformed such that $n-1$ universal blocks are used
  (instead of $n-2$), and the final variable is transformed to the
  overall rapidity of the system, which then defines the momenta of
  the two incoming particles in terms of the final-state ones and this
  variable. See Ref.~\cite{Bothmann:2023siu} for details.
\end{enumerate}

Since the first three methods have the two Bjorken $x$'s as
integration variables, they can also be used for fixed-energy
lepton collisions, by simply setting $x_a=x_b=1$ instead of generating
them. The p$_T$-based integration is thus only evaluated in the ``LHC setup'', described in the Results section.

\subsection{Reweighting and secondary unweighting}
Once the unweighted events following the density of the LC-accurate matrix
elements have been generated using the integration techniques described above, they can be reweighted to improve their
accuracy.  We consider the reweighting of events \cite{Giele:2011cb} both to next-to-leading
colour (NLC) and full-colour accuracy, using the reweight
factors
\begin{equation}\label{eq.rwfactor}
r^{\LC\to\NLC}=\frac{|\M^{\NLC}|^2}{\sum_{i} C_{ii} |\A_i|^2},
\qquad \text{and}\qquad r^{\LC\to\FC}=\frac{|\M|^2}{\sum_{i} C_{ii}
  |\A_i|^2},
\end{equation}
in which $|\M^{\NLC}|^2$ and $|\M|^2$ are the matrix elements
summed over the colours of the external particles, up to
next-to-leading- and full-colour accuracy, respectively. These factors
are phase-space-point dependent and need to be computed for each LC
event separately. Since the external particles in LC events have
assigned helicities, only the helicity configuration chosen for each event needs to be
considered in both the numerator and denominator of the reweight
factors, with no helicity sum required. We remark that this method relies on the fact that the LC matrix-elements are non-zero in all regions of phase space, resulting in finite values of these reweight factors.

Note that the LC events have been generated only in a subset of all the
available channels, i.e., only a subset of the elements in the sum on the
r.h.s.~of Eq.\eqref{eq.LC} need to be taken into account, since the other
contributions can be obtained by swapping identical final-state
particles. However, in the
denominators in Eq.~\eqref{eq.rwfactor} all the elements in the sum need to be
taken into account in the colour ordering to get the correct distribution of
NLC or FC accuracy events.\footnote{We have considered two alternatives to using the colour-summed FC (or NLC)
  and LC matrix elements in Eq.~\eqref{eq.rwfactor}. First, one could take
  only a single row `$i$' in the colour matrix in both the numerator and
  denominator~\cite{Zeppenfeld:1988bz,Frederix:2021wdv}, i.e., the one corresponding to the LC
  colour ordering used to generate the LC unweighted events. While we have
  checked that this does give the correct FC (and NLC) cross section, the
  variance in the reweight factors is enormous, resulting in poor overall
  performance. Moreover, it yields a sizable fraction of negatively weighted
  events, since the numerator is no longer positive definite and is therefore
  not a viable option.  The second alternative is to treat colour similarly to
  helicity; that is, assign to each external particle an explicit
  colour and/or anti-colour, weighted by their relative contribution to the LC
  matrix elements, and compute the numerator and denominator in the reweight
  factor only for that colour assignment. While this is guaranteed to result
  in positive reweight factors, we have found that the variance among these
  factors is much larger compared to summing over all colours, which again
  reduces the performance in the secondary unweighting step.}

The secondary unweighting is performed in the usual manner, in which
a weighted event $i$ is accepted if its reweight factor satisfies
\begin{eqnarray}\label{eq:2ndunw}
  \frac{r_i}{r_{\rm max}} > R\,,
\end{eqnarray}
where $R$ is a uniformly distributed random number between 0 and 1, and $r_{\rm
  max}$ is the maximum possible reweight factor. In practice, the
maximum possible reweight factor is unknown, but a good proxy is the
maximum reweight factor found among all the events that are considered
in the reweighting\footnote{In principle one does not need to
consider the largest reweight factor(s) encountered (and therefore
improve the efficiency because $r_{\rm max}$ will be smaller) as long
as the combined contribution one is neglecting is (well) within the
statistical uncertainty of the event sample.}.

Note that, depending on the accuracy of the LC and NLC approximations
and the relative timing of evaluating the NLC matrix elements
versus the full-colour matrix elements, it might be less computationally
intensive to first reweight and unweight the LC events to NLC
accuracy, and after that reweight them to full-colour using $r^{\NLC\to\FC}=r^{\LC\to\FC}/r^{\LC\to\NLC}$,  and perform a
tertiary unweighting step, since this might significantly reduce the
number of events for which the full-colour matrix elements need to be
evaluated. Since this depends strongly on the evaluation time and
therefore on the details of the implementation of the NLC and
full-colour matrix elements, which we have not fully optimised, we
will only discuss the distribution of the $\textrm{NLC} \to \textrm{FC}$
reweight factors in this work, and leave a full assessment and
viability of this three-step approach to future studies.

\section{Results}\label{sec:results}

\subsection{Setup}

In order to test our approach we consider multi-parton
production in QCD both at a fixed collider energy and with a more realistic setup
including parton luminosities. In both setups we include processes up to 7
partons in the final state. We consider processes with only gluons, processes with one quark
line, and processes with 2 quark lines. For the latter, we consider both same-
and different-flavour quarks, see the first column of Tab.~\ref{tab:proc} for
the complete list of processes considered. For each of these flavour configurations, we consider all
the distinct colour orderings contributing. We have listed the number of them in
columns 2-7 of Tab.~\ref{tab:proc}. As can be seen, the contributing number
of distinct colour orderings is rather modest, even for high-multiplicity
processes. This is because at high multiplicities there are more identical
final-state particles, cancelling the factorial growth in the number of colour
orderings. To be able to investigate the performance of our method in detail,
each possible contribution is kept separate: we do not sum over quark
flavours, nor combine different colour orderings in the integration. We
treat each element in the sum of the r.h.s.~of Eq.~\eqref{eq.LC} as a separate
integral for which we sample the same
number of phase-space points and generate the same number of (unweighted)
events. Note that in practical applications, this is not optimal, since one
would like to compute channels that have a large contribution to the cross
section with a higher accuracy than channels that have a smaller contribution.

\begin{table}[htb!]
\begin{center}
\begin{tabularx}{0.74\textwidth}{l c c c c c c }
\toprule
process & $n=2$ & $n=3$ & $n=4$ & $n=5$ & $n=6$ & $n=7$\\
\midrule
$g g \rightarrow n\, g$ & 2 & 2 & 3 & 3 & 4 & 4\\
\midrule
$g g \rightarrow  d \overline{d} + (n-2) g$ & 1 & 2 & 4 & 6 & 9 & 12\\
$\overline{d} g \rightarrow  \overline{d} + (n-1) g$ & 2 & 3 & 4 & 5 & 6 & 7 \\
$\overline{d} d \rightarrow  n\, g$ & 1 & 1 & 1 & 1 & 1 & 1 \\
\midrule
$g g \rightarrow  d \overline{d} d \overline{d} + (n-4) g$ & -& -& 2 & 4 & 11 & 18 \\
$d g \rightarrow  d d \overline{d} + (n-3) g$ & -& 2 & 6 & 12 & 20 & 30\\
$d \overline{d} \rightarrow  d \overline{d} + (n-2) g$ & 2 & 3 & 5 & 6 & 8 & 9 \\
$d d \rightarrow  d d + (n-2) g$ & 1 & 1 & 2 & 2 & 3 & 3 \\
\midrule
$g g \rightarrow  d \overline{d} u \overline{u} + (n-4) g$ & -& -& 4 & 8 & 22 & 36 \\
$d g \rightarrow  d u \overline{u} + (n-3) g$ & -& 4 & 12 & 24 & 40 & 60\\
$d \overline{d} \rightarrow u \overline{u} + (n-2) g$ & 2 & 3 & 5 & 6 & 8 & 9\\
$d u \rightarrow  d u + (n-2) g$ & 2 & 2 & 4 & 4 & 6 & 6\\
\bottomrule
\label{tab:proc}
\end{tabularx}
\end{center}
\caption{Number of distinct elements (``channels'') contributing to the sum in
  the r.h.s.~of Eq.~\eqref{eq.LC}.}
\end{table}

In each of the channels, we use 640$\times 10^3$ phase-space points (that pass the
generation cuts) to setup the importance-sampling grids. After this, the grids
are kept fixed and 6.4$\times10^6$ points (passing the generation cuts) are used to
compute the phase-space integral, and determine the maximum weight for the (primary)
unweighting. Instead of using a single maximum weight, we follow the procedure
outlined in Ref.~\cite{Nason:2007vt} to form a phase-space dependent upper
bounding envelope\footnote{We have made various small changes to the method
  introduced in Ref.~\cite{Nason:2007vt}, of which the most important one is a
  more aggressive increase in the upper bound: following the notation of
  Ref.~\cite{Nason:2007vt}, if $\bar{f}>\prod_{k=1}^n u^k(z_k)$, we increase
  each $u^k(z_k)$ by
  \[
    f=\textrm{min}\left(\frac{\bar{f}}{\prod_{k=1}^n u^k(z_k)},2\right)^{1/n}.
  \]
  This results in a faster convergence to a stable upper bound.}. To assess
the efficiency of the integration methods, for each of the channels, we
investigate the relative Monte Carlo uncertainties of the cross section. More importantly, we also investigate the unweighting efficiency for the generation of events,
defined as
\begin{eqnarray}
  \epsilon_1 = \frac{N}{\tilde{N}},
\end{eqnarray}
where $N=100$ is the number of unweighted events generated and
$\tilde{N}$ is the total number of phase-space points (that pass the
cuts) tried to generate these $N$ unweighted events. For each of the channels, this
procedure is repeated $4\times 10$ times for each of the 4 distinct
phase-space parametrisations and for 10 different random number seeds. The
latter allows us to estimate the stability of the results and assess the
(potential) sensitivity to statistical outliers.

To study the secondary unweighting efficiency, computed as the normalised sum
of all the ratios of the weights
\begin{eqnarray}\label{eq.2unw}
  \epsilon_{2} = \frac{1}{N}\sum_{i=1}^N  \frac{r_i}{r_{\rm max}},
\end{eqnarray}
we need larger event samples. Therefore, in each channel and for each of the
10 different random number seeds, we generate $N=100\times10^3$ unweighted
events. This is done for only one of the integration methods, since the four
phase-space parametrisations yield statistically equivalent event files. We
reweight the unweighted events both to $\textrm{LC} \to \textrm{FC}$ and
$\textrm{LC} \to \textrm{NLC}$, with the latter subsequently reweighted to
$\textrm{NLC} \to \textrm{FC}$ accuracy\footnote{The
  $\textrm{NLC} \to \textrm{FC}$ efficiency does not take into account the
  change in weight distribution after the $\textrm{LC}\to \textrm{NLC}$
  reweighting and unweighting, since we compute the reweight factor for
  \emph{each} LC event generated, rather than a subset corresponding to the
  unweighted NLC events.}.

A few comments on the definition of LC and NLC accuracy are in order. Firstly, as
already discussed in Sec.~\ref{sec.LC}, in the case of the different-flavour
two-quark line processes, the definition of LC is not strictly that of a
leading-colour approximation. In particular, all the (distinct) elements on
the diagonal of the colour matrix, defined in the fundamental
representation, are included as separate channels. This includes terms that
contribute only at NLC accuracy. Secondly, for the NLC approximation, for the all-gluon
processes, we strictly truncate all contributions beyond the NLC
accuracy. Hence they include all the terms which are proportional to $\NC^{n+2}$
and $\NC^n$ and nothing beyond
that. On the other hand, for the processes with at least one quark line, we do
not truncate all the contributions beyond NLC. That is, we perform the colour
decomposition of the amplitudes using the fundamental basis, and as soon as a
coefficient in the colour matrix contributes at NLC, we include the
full-colour version of that coefficient in our calculation. We have found that for these processes
this yields a NLC approximation closer to the FC result as compared to
truncating the value of the coefficient to NLC accuracy.

In this work we consider two setups for the event generation:
\begin{description}
\item[Fixed partonic collision energy]
In this setup, we use a fixed partonic collision energy of
$\sqrt{s}=1000~\textrm{GeV}$, without the inclusion of initial-state parton
distribution functions. We apply generation cuts on the two-body
invariants $|2 p_i\cdot p_j| \ge 900~\textrm{GeV}^2$, where $i \ne j$,
including both initial- and final-state particles, and for the strong
coupling we use a fixed value of $\alpha_s=0.119$.
\item[LHC collision at 14~TeV]
In this more realistic setup, we consider LHC collisions at $14~\textrm{TeV}$,
using the NNPDF2.3NLO PDF set~\cite{Ball:2012cx}. The value of the strong coupling is
still kept fixed at $\alpha_s=0.119$, and the factorisation scale is set equal
to the $Z$-boson mass. As generation cuts, we select events that have all
final-state partons at a minimum transverse momentum of
$p_T > 30~\textrm{GeV}$, a maximum (absolute) rapidity of $|y| < 6.0$ and a
minimum separation between them of $\Delta R > 0.4$.
\end{description}

Except for the largest multiplicities considered, we have checked that the
cross sections obtained with the setups above agree within statistical
uncertainties with known results.

\subsection{Numerical results for fixed partonic collision energy}
\begin{figure}[htb!]
  \begin{center}
    \includegraphics[width=0.48\textwidth]{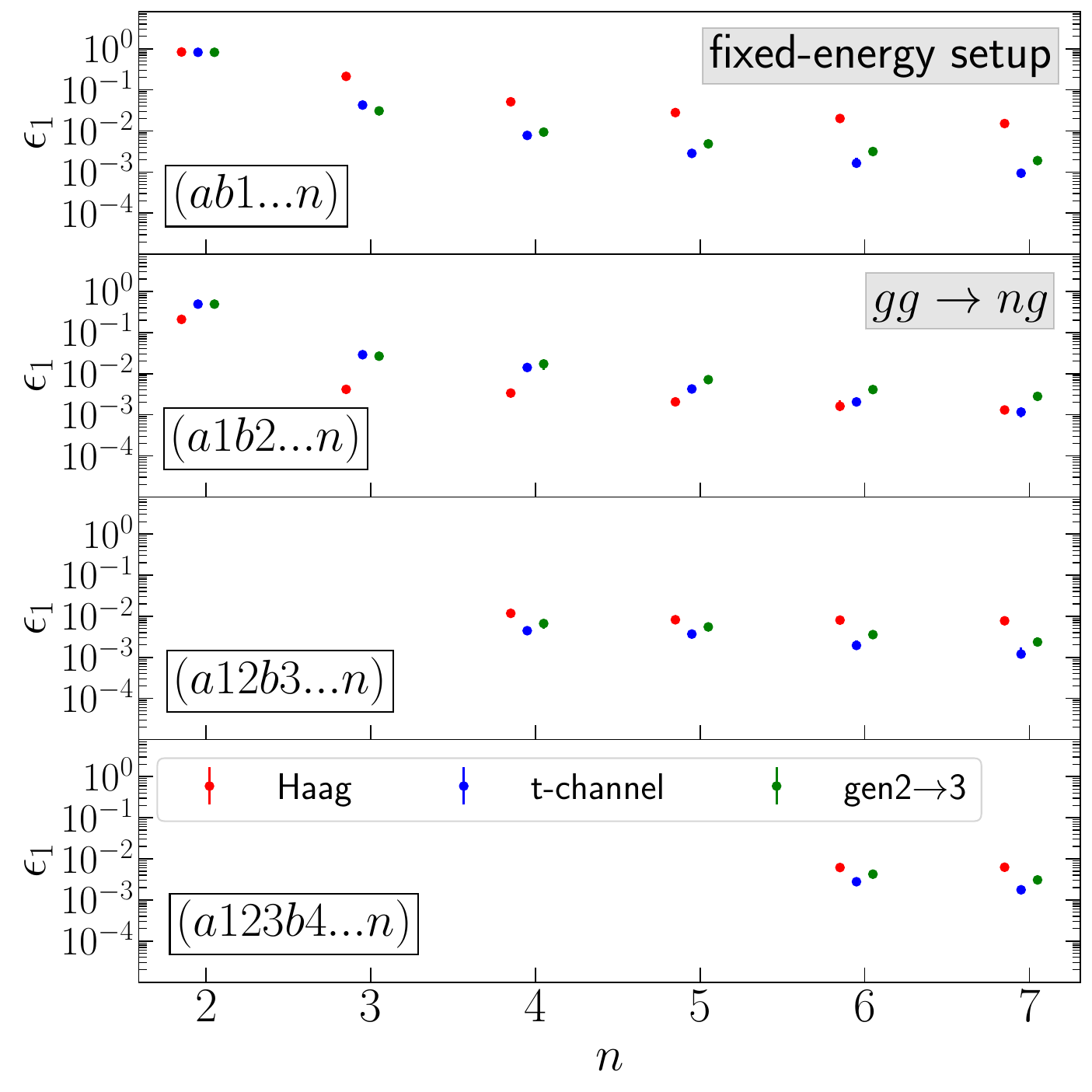}
    \includegraphics[width=0.48\textwidth]{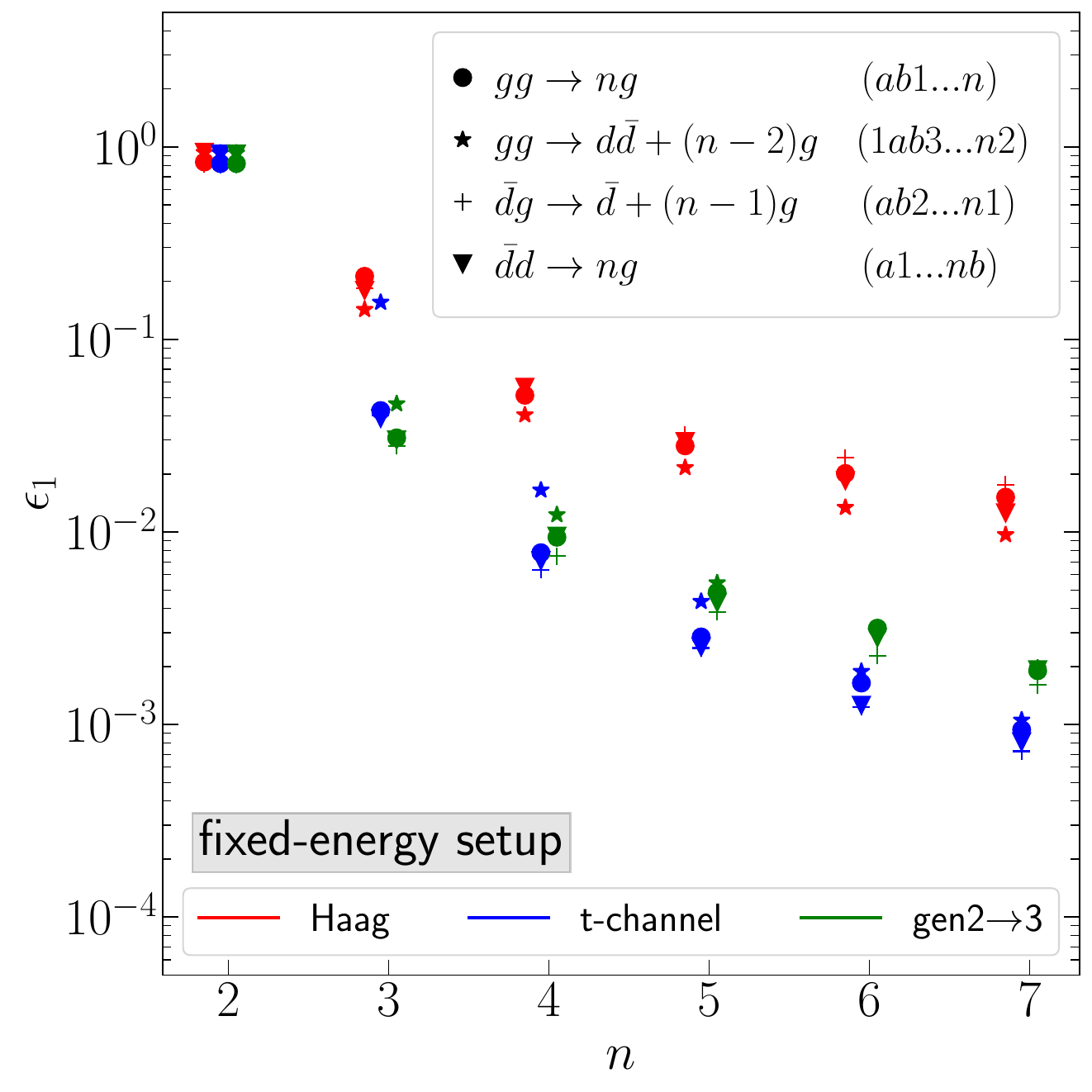}
    \caption{Unweighting efficiency, $\epsilon_1$, for the phase-space
      integration for the $gg\to n\,g$ process (left plot), using the
      LC approximation of the integrand, for
   $n=2$ to $n=7$ number of final-state partons, the three integration methods and for
      all the non-equivalent colour orderings. In the right-hand plot we
      show $\epsilon_1$ also for the process with one $q\bar{q}$
      pair, for the colour ordering in which all final-state gluons
      are adjacent. Both plots show the fixed partonic collision energy setup. }
    \label{fig:FCE_unw_eff}
  \end{center}
\end{figure}

In the left-hand plot of Fig.~\ref{fig:FCE_unw_eff} we present the
unweighting efficiencies, $\epsilon_1$, for the generation of LC events
for the all-gluon processes. Processes with 2 up to 7 final-state gluons
are considered. The four separate insets correspond to the (up to)
four distinct colour orderings that need to be considered at LC accuracy,
with the explicit colour ordering expressed in a string in the insets,
using $a$ and $b$ for the labels of the
right- and left-incoming partons and $1, 2, \ldots, n$ for the labels of the
$n$ outgoing partons. The various colours correspond to the phase-space
parametrisations used: red for the Haag method, blue for the
t-channel, and green for the \gen~parametrisation\footnote{As discussed
in Sec.~\ref{sec.PS}, the p$_T$-based generator cannot be used for
collisions at fixed partonic energy.}. Since we have considered 10
different random seeds for each setup, we include vertical bars through
the dots, showing the (barely visible) envelope of the 10 unweighting efficiencies obtained.

As expected, irrespective of the colour ordering and the phase-space
parametrisation, the unweighting efficiency is much higher at low
multiplicities compared to when more particles are produced in the final
state. The differences between the colour orderings are not
significant for the t-channel and the \gen\ parametrisations, being
above 50\% for $gg \to 2 g$ and decreasing to the (several) per-mille
level for $gg \to 7 g$. With the exception of the $n=2$ and $n=3$
multiplicities, the \gen\ method yields slightly larger efficiencies
than the t-channel parametrisation. On the other hand, the Haag method
behaves quite differently for each of the colour orderings. This method
is designed to generate the phase-space density according to the
all-gluon MHV matrix elements, and works particularly well for the
$(ab12\ldots n)$ colour ordering, where all the final-state gluons are
adjacent in the colour ordering. Indeed, for that case,
Haag outperforms the other two methods by about an order of
magnitude. For the $(a1b2\ldots n)$, i.e., where one of the final-state gluons
is separated from the others, the behaviour is
worse: the efficiency of the Haag method is significantly lower than
the other two parametrisations, in particular for moderate
multiplicities. For the other two distinct colour orderings, Haag is
slightly more efficient than the t-channel and \gen\ methods, but still not as
good as for the $(ab12\ldots n)$ ordering. The reason why the
$(ab12\ldots n)$ colour ordering works most efficient with Haag is that
for this configuration, all the final-state partons are constructed
with antenna blocks~\cite{vanHameren:2002tc}, while in
particular for $(a1b2\ldots n)$, the momentum of gluon $1$ is not
generated according to an antenna block.

Similarly to what can be observed for the all-gluon processes in the
left-hand plot of Fig.~\ref{fig:FCE_unw_eff}, we can see in the
right-hand plot for the processes including one quark line. In the
latter plot, we only consider the colour ordering where all final-state gluons are adjacent, and thus Haag is
expected to be most optimal. Indeed, this is also what can be
observed. Haag outperforms the other two phase-space parametrisations for all process types
by a rather significant margin. However, we have checked that this is
not the case for most of the other colour orderings one must consider,
and the behaviour in this respect for the one-quark-line processes is very similar to
the all-gluon process, discussed above.

\begin{figure}[htb!]
  \begin{center}
    \includegraphics[width=0.48\textwidth]{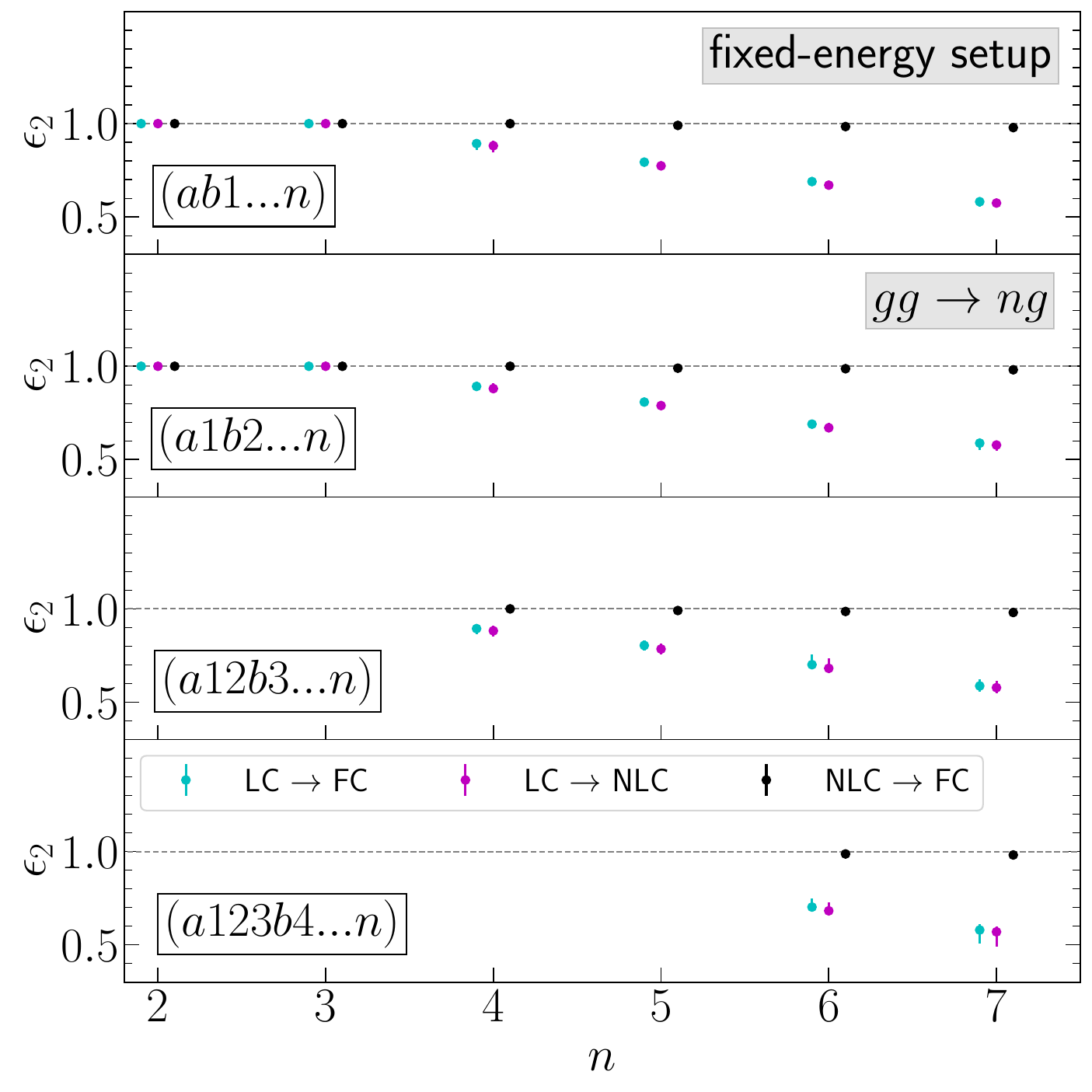}
    \caption{Secondary and tertiary unweighting efficiencies, $\epsilon_2$, for the
      all-gluon processes, $gg\to n\,g$, obtained after reweighting the
      LC to NLC and to FC, and the NLC to FC as a function of the gluon multiplicity $n$, for the fixed partonic energy setup.}
    \label{fig:FCE_2nd_unw_eff}
  \end{center}
\end{figure}

In Fig.~\ref{fig:FCE_2nd_unw_eff} we present the secondary (and
tertiary) unweighting efficiencies, $\epsilon_2$, for the $gg \to n\,
g$ process as a function of $n$. Similarly to the left-hand plot of
Fig.~\ref{fig:FCE_unw_eff}, we consider all the (up to) four
distinct colour orderings. In cyan dots, we show the efficiency for
the $\textrm{LC} \to \textrm{FC}$ reweighting, in magenta the
$\textrm{LC} \to \textrm{NLC}$ efficiency and in black the (tertiary)
$\textrm{NLC} \to \textrm{FC}$ unweighting efficiency, and, as before,
the vertical bars cover the extrema among the
unweighting efficiencies obtained from the 10 independent samples
produced.

As expected, the $\epsilon_2$ measure is exactly equal to one for $n=2$ and $n=3$ in
the all-gluon processes, because the beyond-LC contributions only result in an
overall shift by the factor $(1-1/\NC^2)$. Starting from $n=4$, this is no longer the case
and both reweighting factors $r^{\LC\to\NLC}$ and $r^{\LC\to\FC}$, as well as
the tertiary reweighting factor $r^{\NLC\to\FC}$ gain dependence on the
kinematics of the event being reweighted. Increasing the multiplicity to
$n=7$, the variance in the factors increases, resulting in reduced
efficiencies, but at all multiplicities considered, the secondary unweighting efficiencies
$\LC\to\NLC$ and $\LC\to\FC$ remain above 50\%, for all
colour orderings. It should not come as a surprise that there is no large
dependence on the colour ordering here: in the denominator of the reweight
factors, Eq.~\ref{eq.rwfactor}, all colour orderings are summed over. Hence,
the only dependence on the colour ordering comes through the kinematics of the
generated LC event.  In general, the $\LC\to\NLC$ reweighting results in a
slightly larger efficiency than the direct $\LC\to\FC$ reweighting.
And, the unweighting efficiency for $\NLC\to\FC$ is always greater
than 95\%. This is a positive sign that an efficient three-step approach is possible:
(1) generate LC
unweighted events, (2) reweight and unweight those events to NLC
accuracy, and (3) only then reweight (and unweight) the remaining events to FC
accuracy. The outcome of this approach depends on the details of the implementation: not only the
efficiencies, but also the 
timing (and potentially memory usage) of the LC, NLC and FC matrix elements 
determine if this three-step strategy is viable in practice. Since our
implementations are not fully optimised, we refrain from making any claims
here and leave it for future studies.

From this simplified setup, i.e., fixed partonic collision energy and
generation cuts on all the two-particle invariants, we can conclude
that our approach of generating unweighted LC events first, and only
then reweighting them (and unweighting again) to FC (or NLC) accuracy,
is viable. However, this setup is very much an ideal case. In
particular, the Haag integrator is known to behave rather well when
cuts are applied on the two-body invariants and not on the more
commonly used kinematical variables at LHC, such as transverse momentum, rapidity
and angular separation. We will study the latter in more detail in the
next section.

\subsection{Numerical results for the LHC setup}

\subsubsection*{Generation of LC events}
\begin{figure}[htb!]
  \begin{center}
    \includegraphics[width=0.48\textwidth]{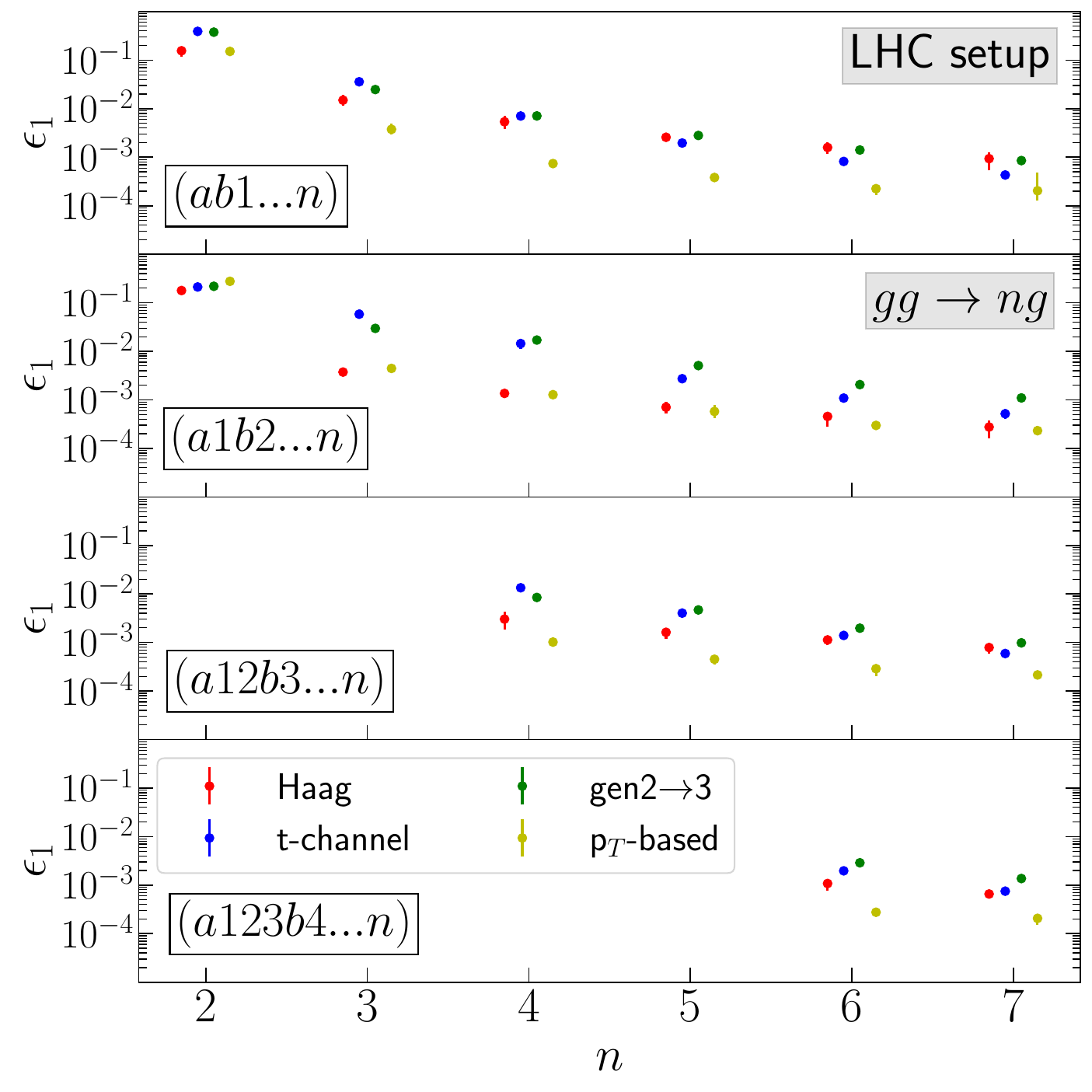}
    \includegraphics[width=0.48\textwidth]{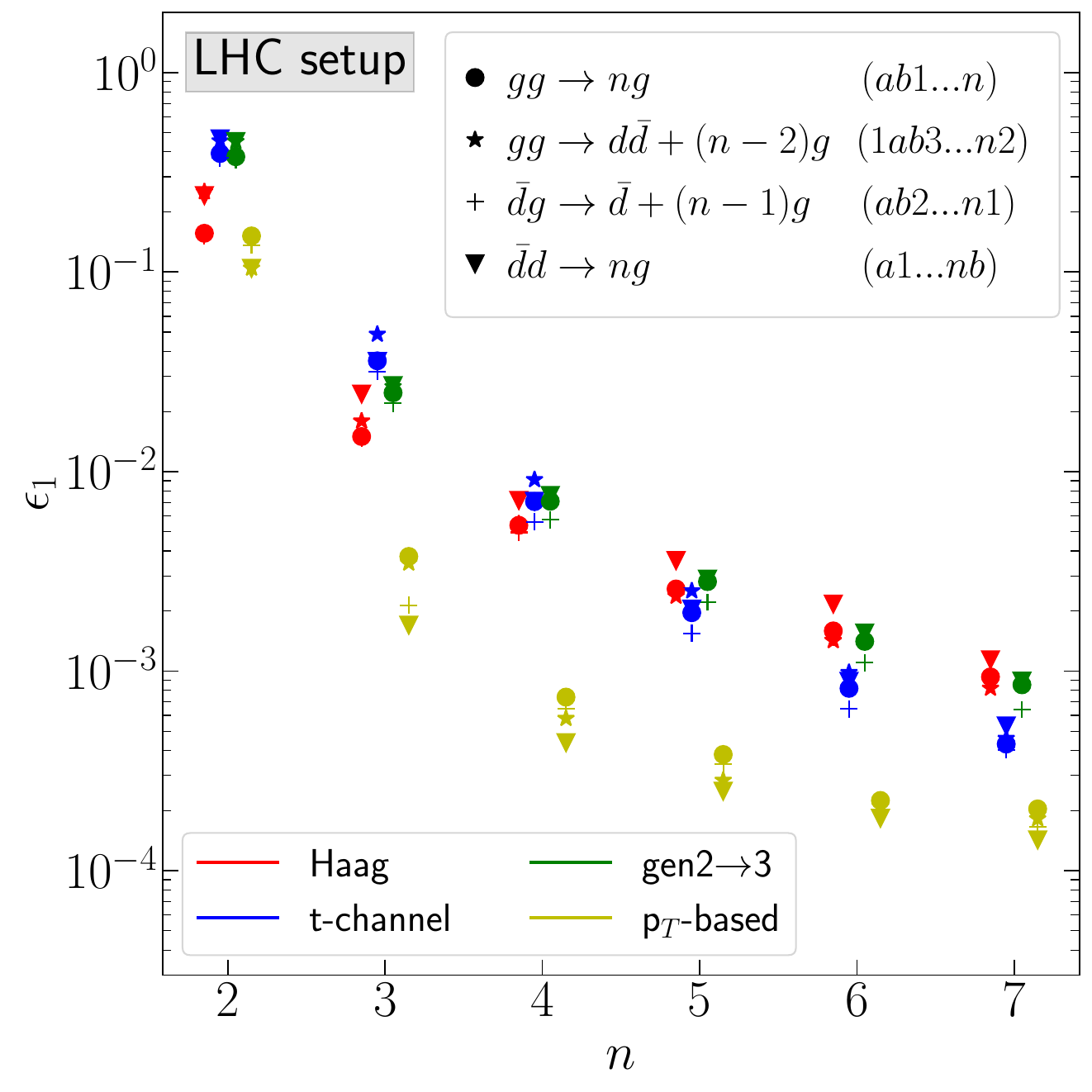}
    \caption{Unweighting efficiency, $\epsilon_1$, for the phase-space
      integration for the $gg\to n\,g$ processes (left plot), using the
      LC approximation of the integrand, for the number of final-state
      partons $n=2$ to $n=7$, the four integration methods and for
      all the non-equivalent colour orderings. In the right-hand plot we
      show the $\epsilon_1$ measure also for the processes with one $q\bar{q}$
      pair, for the colour ordering in which all final-state gluons
      are adjacent.  Both plots show the LHC setup. }
    \label{fig:unw_eff}
  \end{center}
\end{figure}

In this more realistic setup with included parton luminosities, we can see in Fig.~\ref{fig:unw_eff}, that the overall behaviour of
the unweighting efficiencies in generating the LC events is rather similar to the fixed partonic
collision energy results. In this figure, where we present exactly the
same results as in Fig.~\ref{fig:FCE_unw_eff}, but for the LHC setup,
we have now also included the results for the p$_T$-based phase-space
parametrisation in yellow. For all the results presented here,
this method has a worse performance than the other three
parametrisations. This is not surprising: this method does not have
any integration variables directly aligned with the peaks in the LC
matrix elements, resulting in a poorer adaptation of the importance
sampling towards the integrand. However, this is rather specific to integrating
single colour orderings at a time. If one
would integrate directly FC matrix elements (or multiple colour
orderings together) the advantages of using Haag, t-channel or
\gen\ parametrisation are greatly reduced, while the p$_T$-based one would
be less affected.

By comparing the unweighting efficiencies for the LHC setup (Fig.~\ref{fig:unw_eff}) with the ones
for the fixed partonic collision energy
(Fig.~\ref{fig:FCE_unw_eff}), we notice a slightly poorer
overall performance. This comes as no surprise, since there are two extra
integration variables (the Bjorken $x$'s) and slightly more
complicated phase-space cuts in the LHC setup, the variance among the phase-space
points increases somewhat, resulting in the decreased efficiency. This
is particularly noticeable for the Haag parametrisation, that performs
considerably worse with cuts on the $p_T$, $\eta$ and $\Delta R$ as
compared to the two-body invariants in the fixed partonic collision energy setup.
The main reason is that these cuts impact the integration boundaries
of the variables used in the Haag integration in a complex manner, meaning that
optimisation is difficult, resulting in many generated phase-space
points not passing these cuts. Even though these points are not
included in the computation of the unweighting efficiencies, they do
impact the setup of the importance sampling grids, potentially
increasing the variance among the points passing the cuts.

\begin{figure}[htb!]
  \begin{center}
    \includegraphics[width=0.8\textwidth]{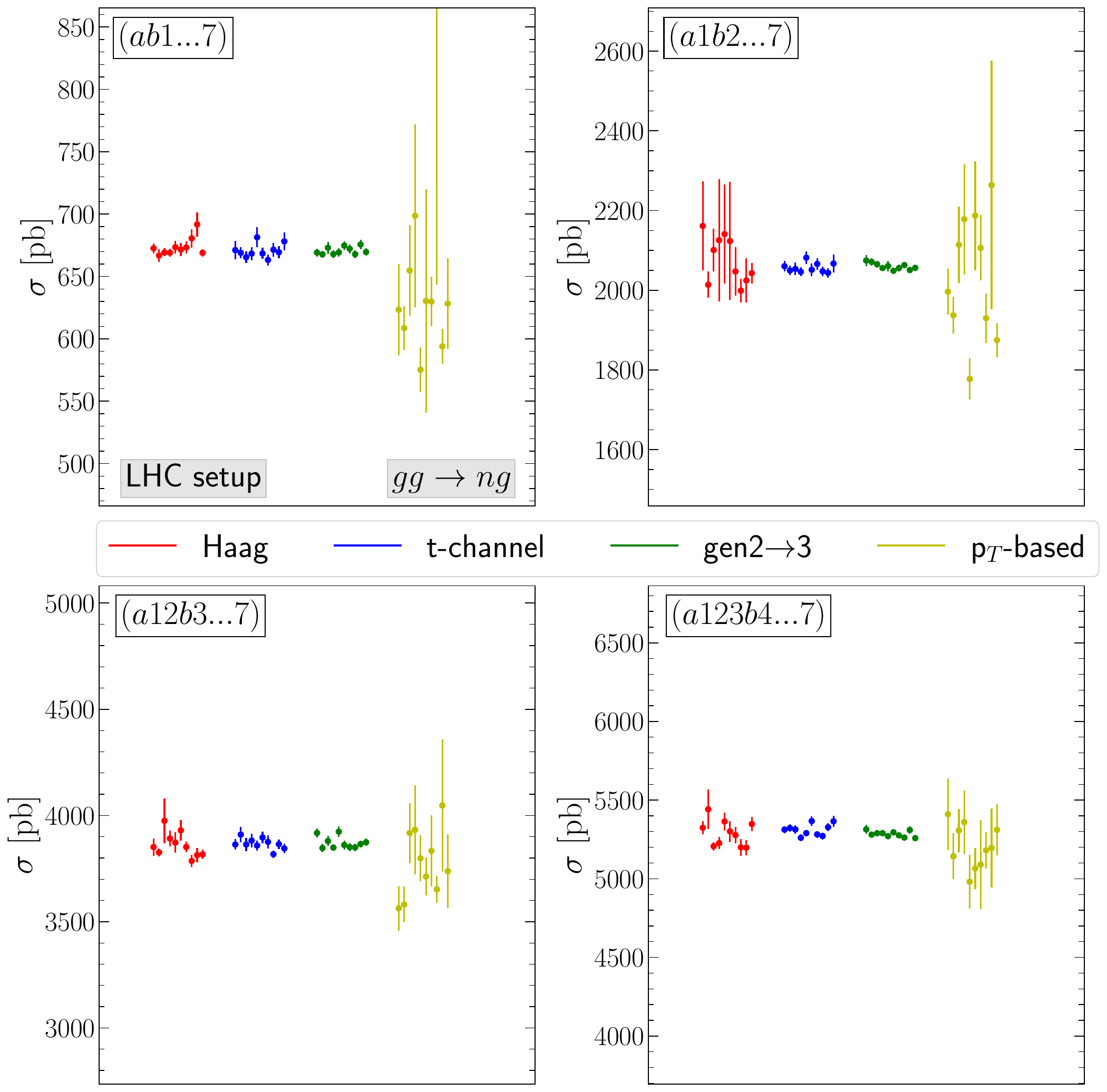}
    \caption{Cross section (in pb) for the $gg \to 7g$ process for the
      four distinct colour orderings and the four
      different phase-space parametrisations for the LHC setup. Each setup is repeated
      10 times with a different random number seed, resulting in the
      10 points per integration method. The vertical bars indicate the statistical
      Monte Carlo uncertainty on the cross sections. The cross sections
      include a factor $7!$, corresponding to the final-state symmetry factor.}
    \label{fig:MC_unc}
  \end{center}
\end{figure}

The values for the unweighting efficiencies are highly correlated with
the statistical uncertainties obtained from the Monte Carlo
phase-space integration, presented in Fig.~\ref{fig:MC_unc}. In this figure,
the LC cross section (in units of pb) for the $gg \to 7g$ process is
shown. Results are separated in the four distinct colour
orderings, and repeated for the four phase-space
parametrisations considered. Moreover, each complete setup is repeated
10 times with different random number seeds.
As can be seen, all the obtained cross sections among the $4 \times 10$ runs are
compatible within their statistical uncertainties (the $1\sigma$
statistical errors are given by the vertical bars). Moreover, the
uncertainties obtained and the spread among the cross sections are
the smallest with the \gen\ parametrisation. This is compatible with
the fact that this setup has the largest unweighting
efficiency. Similarly, the p$_T$-based method shows the largest
spread and uncertainties in the cross section, resulting
in the smallest unweighting efficiency, as discussed above.

\begin{figure}[htb!]
  \begin{center}
    \includegraphics[width=0.48\textwidth]{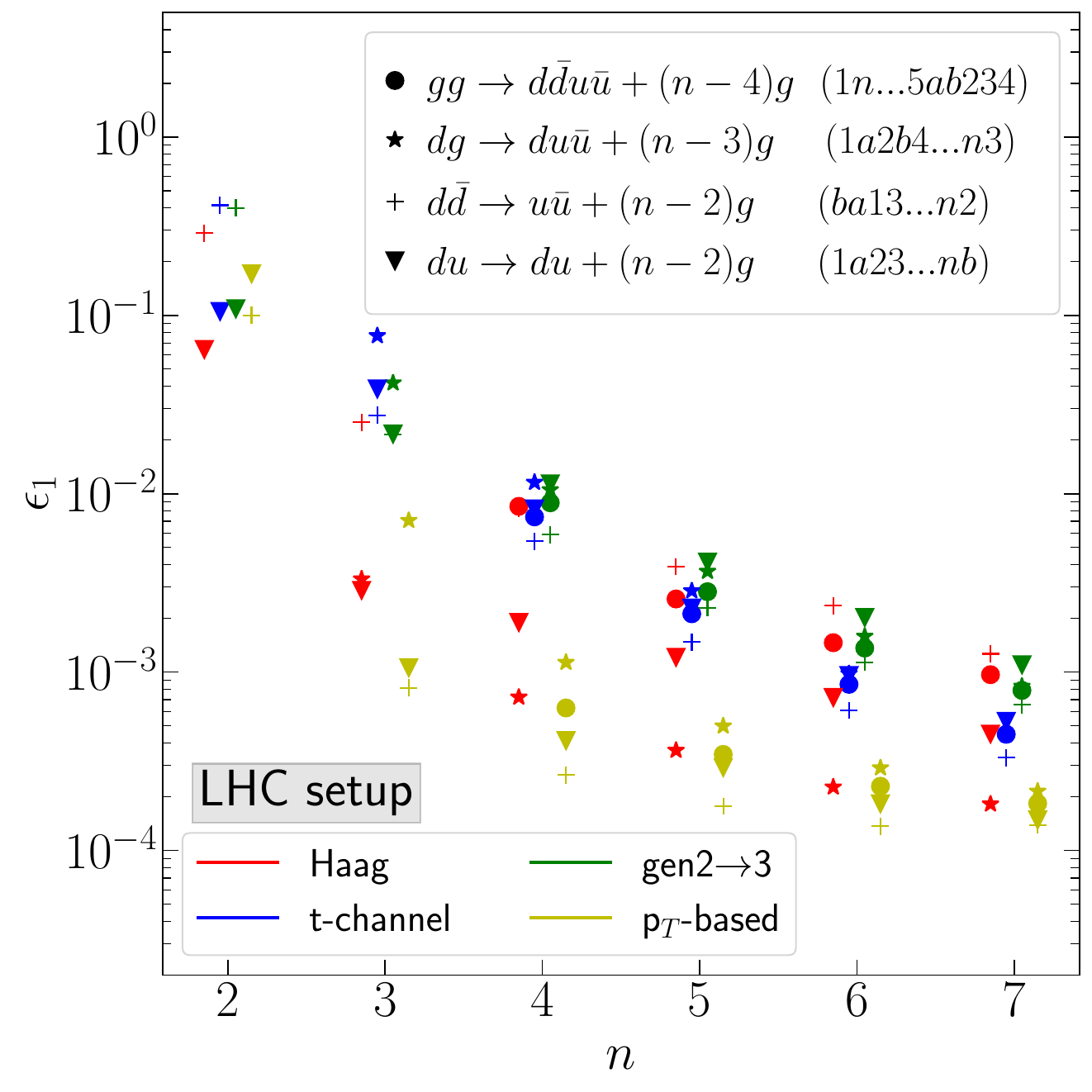}
    \includegraphics[width=0.48\textwidth]{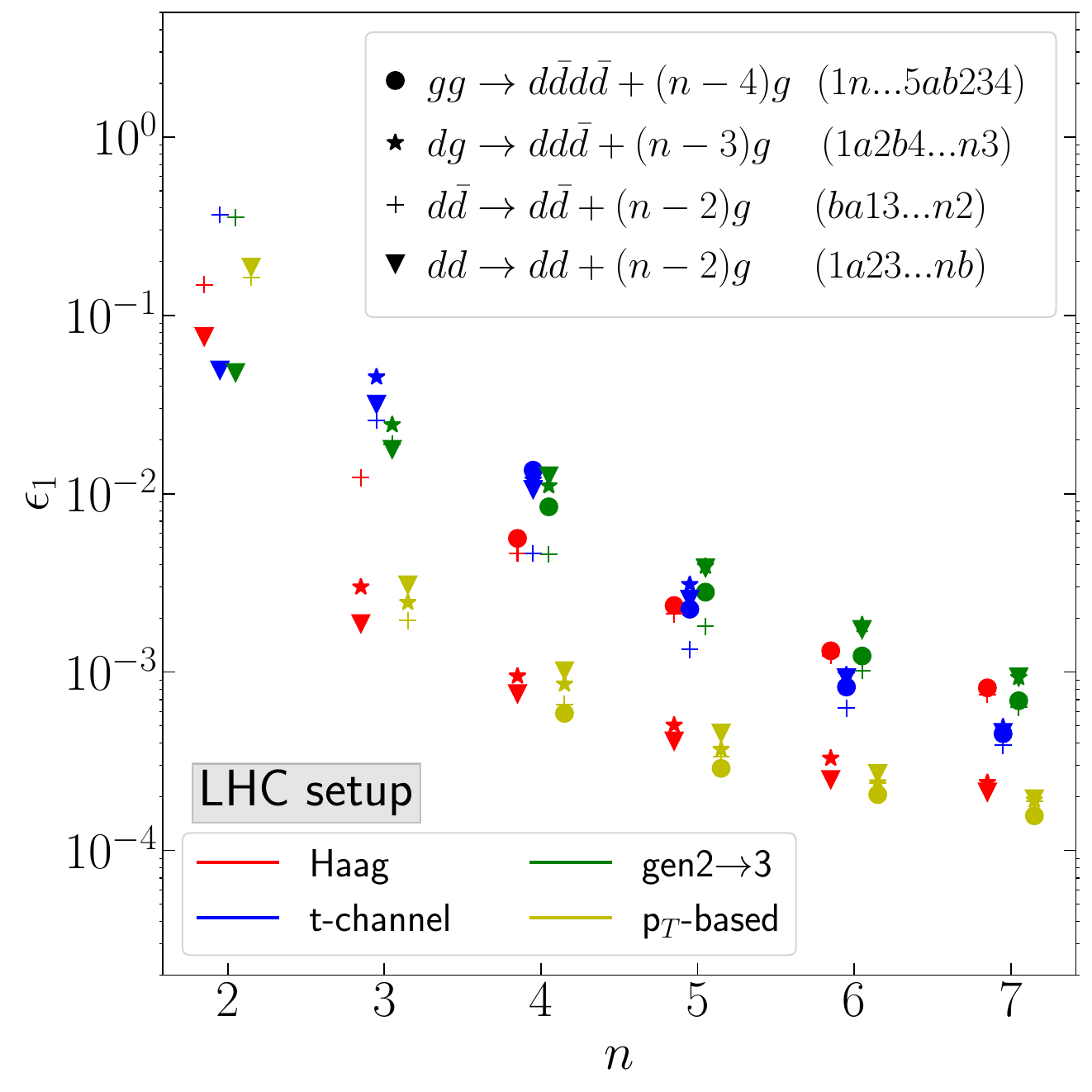}
    \caption{Unweighting efficiency, $\epsilon_1$, for the phase-space
      integration for the two quark-pair process with different
      flavours (left plot) and same flavours (right plot) for the LHC setup, using the
      LC approximation of the integrand, for the number of final-state
      partons $n=2$ to $n=7$, using the four integration methods. Only a
      single colour ordering is considered for each of the flavour
      channels, specified in the legend.}
    \label{fig:unw_eff_2qq}
  \end{center}
\end{figure}

The main features of the results for the unweighting efficiencies for the generation of the
LC events obtained for the $gg\to n\,g$, $g g \rightarrow d
\overline{d} + (n-2) g$, $\overline{d} g \rightarrow \overline{d} +
(n-1) g$, and $\overline{d} d \rightarrow n\, g$ processes translate
to the processes with two quark lines. In Fig.~\ref{fig:unw_eff_2qq}
we present these values for the different-flavour (left plot) and
same-flavour (right plot) quark lines. In these plots, only one of
the colour orderings is considered (one of orderings with most of the final-state gluons adjacent), but the results for the
t-channel, \gen, and p$_T$-based parametrisations are representative
for the other colour orderings. For the Haag method, however, similarly to the all-gluon
process, see left-hand plot of Fig.~\ref{fig:unw_eff}, the colour
ordering with all final-state gluons adjacent works best, and for the other orderings (not-shown), Haag performs somewhat worse compared to the other
methods.

From the results in these figures, we can conclude that, overall, the \gen\
method behaves excellently (with the t-channel a close second) for all processes, colour orderings and
multiplicities, with unweighting efficiencies at (or above) the per-mille
level, even for $2\to 7$ processes.

\subsubsection*{Reweighting the LC events}

\begin{figure}[htb!]
  \begin{center}
    \includegraphics[width=\textwidth]{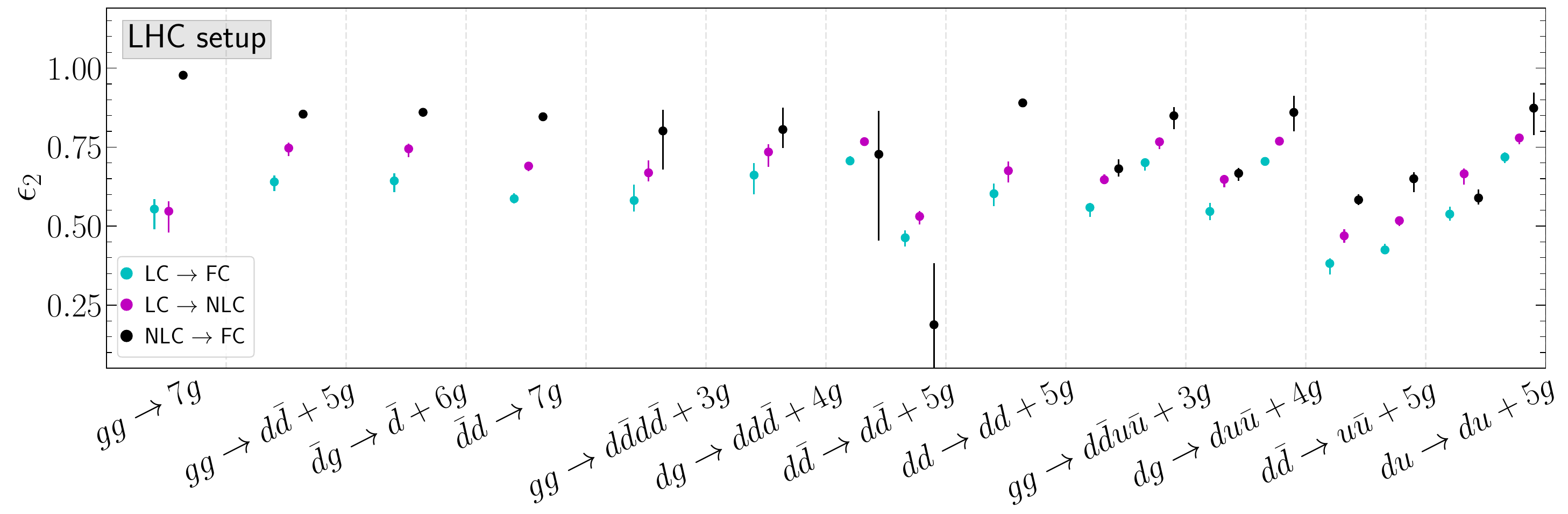}
    \caption{Secondary (and tertiary) unweighting efficiencies, $\epsilon_2$, for all the
      $2 \to 7$ processes for various flavours, for one or two representative colour orderings, for the LHC setup. }
    \label{fig:2nd_unw_eff_all_procs}
  \end{center}
\end{figure}

The main results for the secondary (and tertiary) unweighting efficiencies,
$\epsilon_2$, are summarised in Fig.~\ref{fig:2nd_unw_eff_all_procs}. In
this figure, we have focused on the highest multiplicity that we have
considered, $n=7$, and, for each process, have selected one or two representative
colour orderings.
In this figure, $\epsilon_2$ is represented by the
cyan, magenta and black dots for the reweighting of $\LC\to\FC$,
$\LC\to\NLC$, and $\NLC\to\FC$, respectively.

\begin{figure}[htb!]
  \begin{center}
    \includegraphics[width=0.48\textwidth]{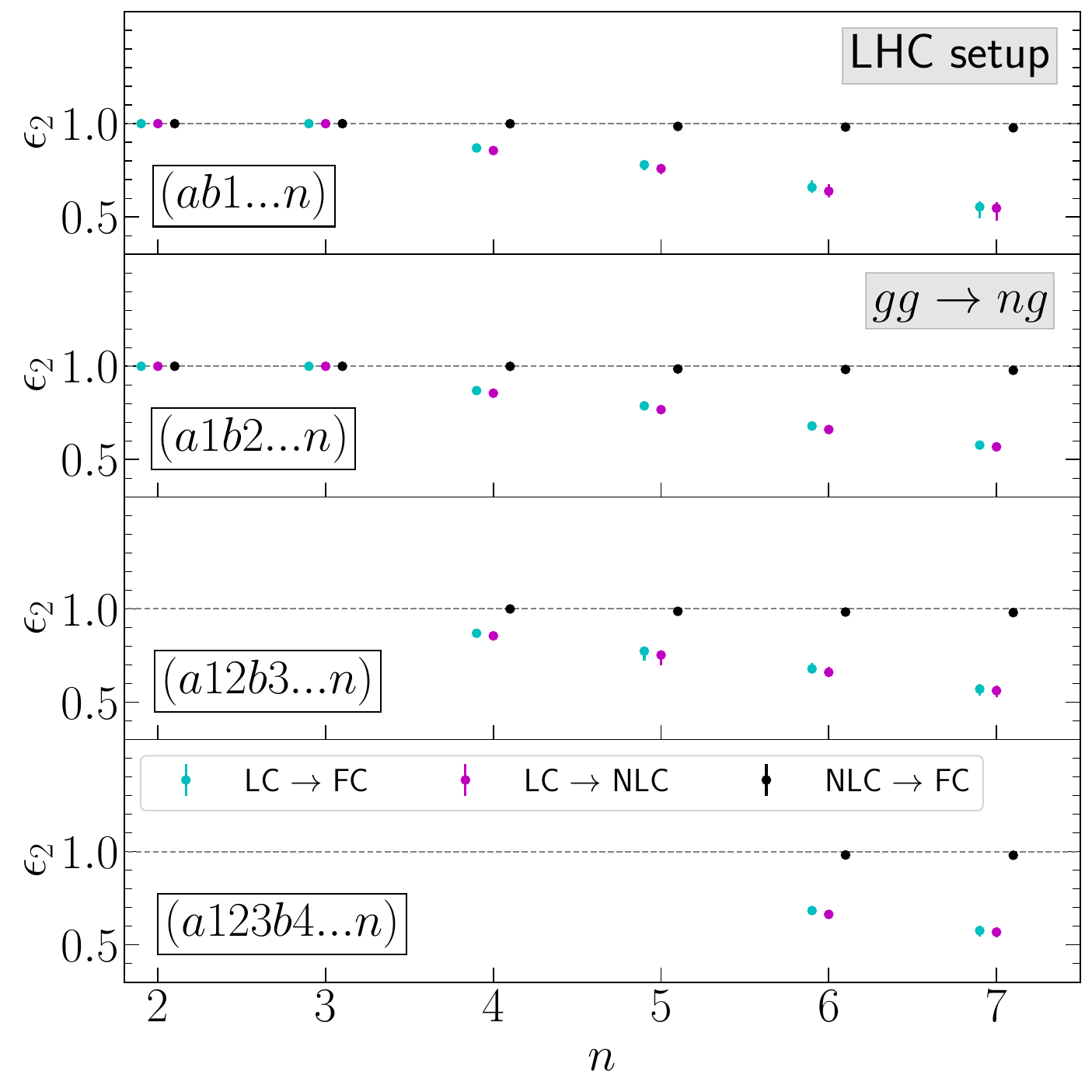}
    \caption{Secondary (and tertiary) unweighting efficiencies, $\epsilon_2$, for the
      reweighting of the LC events for the $gg\to n\,g$ process, for $n=2$ to $n=7$ number of final-state partons, and for all the
      non-equivalent colour orderings, for the LHC setup.}
    \label{fig:2nd_unw_eff}
  \end{center}
\end{figure}

In the first column of this figure, the 
$gg\to 7g$ process is shown. The unweighting
efficiencies $\LC\to\FC$ and $\LC\to\NLC$ are about 50\% for this channel,
with some minor
variations for the 10 different random seeds, represented by the small
vertical bars on the dots. The tertiary unweighting efficiency coming from the
$\NLC\to\FC$ reweighting is rather large, above $95\%$. The secondary (and tertiary) unweighting efficiencies, $\epsilon_2$, for the other multiplicities
and colour orderings for the all-gluon processes are shown in Fig.~\ref{fig:2nd_unw_eff}. We find similar
results as for the fixed partonic collision energy
(Fig.~\ref{fig:FCE_2nd_unw_eff}). Even though the kinematics of the events
in the fixed partonic collision energy setup differ from those in
the LHC setup, the reweighting factors are similar and the secondary (and
tertiary) unweighting efficiencies are very close in value.

\begin{figure}[htb!]
  \begin{center}
    \includegraphics[width=\textwidth]{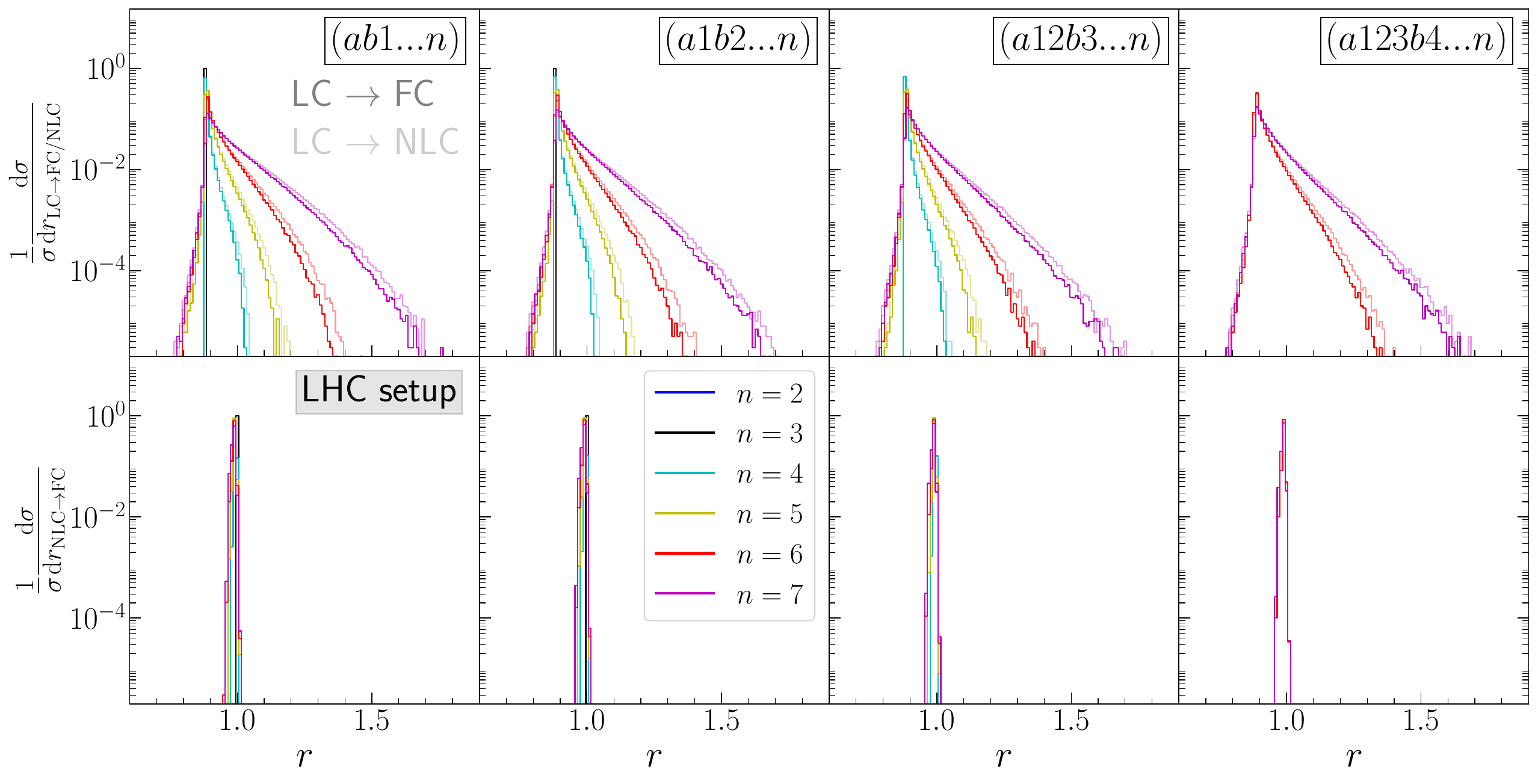}
    \caption{Weight distributions for the reweight factors $r^{\LC\to\FC}$
      and $r^{\LC\to\NLC}$ in the top row and $r^{\NLC\to\FC}$ in the bottom
      row for the all-gluon processes, for the various distinct colour orderings in the four columns, for the LHC setup. The colours represent the different final-state multiplicities.}
    \label{fig:2nd_unw_eff_wd}
  \end{center}
\end{figure}

In Fig.~\ref{fig:2nd_unw_eff_wd} we show the actual weight distributions
corresponding to these unweighting efficiencies for the all-gluon processes. In the top row, the weight
distributions of the $r^{\LC\to\FC}$ and $r^{\LC\to\NLC}$ reweight factors are
shown in the darker and lighter lines, respectively, and in the bottom row the
tertiary $r^{\NLC\to\FC}$ reweight factor. The colours represent the different final-state multiplicities that we consider, and the colour
ordering is specified in the subplots. For these plots, and any following plots
that show the specific weight distributions, we group the events created with
the 10 different random seeds together in one event file. Hence, each of the
distributions shown in these figures are generated from 10$^6$ unweighted
LC events. 

As expected, the $n=2$ (blue) and $n=3$ (black) distributions are single
peaks, located at $r^{\LC\to\FC}=r^{\LC\to\NLC}=(\NC^2-1)/\NC^2$, and
$r^{\NLC\to\FC}=1$, since the NLC contributions are an overall shift of the LC
results for the four- and five-gluon amplitudes. For $n=4$ and above, the reweight
factors are no longer constant over the phase space. Moreover, there are
beyond-$\NLC$ contributions, such that $r^{\LC\to\FC}\ne
r^{\LC\to\NLC}$ in general. However, the two curves remain close to each other for all the
multiplicities and colour orderings considered. Furthermore, these weight
distributions remain very much peaked at $(\NC^2-1)/\NC^2$, with a somewhat
increasing tail towards larger reweight factors at higher
multiplicities. However, even for $n=7$, they remain below about a factor two
above the peak, which corresponds to the about 50\% secondary unweighting
efficiencies found for this multiplicity as depicted (by the cyan and magenta
dots) in Fig.~\ref{fig:2nd_unw_eff}. Note also that by comparing the weight
distributions for the up-to-four distinct colour orderings, they behave rather
similarly. This is because the colour ordering itself does not enter the
actual reweight factors directly; only through the kinematics of the
unweighted LC events, the colour ordering plays a role in the reweighting. The
very large tertiary unweighting efficiencies, $>95\%$, correspond to the very
narrow peaks in the $\NLC\to\FC$ weight distributions (lower row). They are
peaked exactly at 1 for the low to medium multiplicities, but shift to a few
percent below 1 for highest multiplicities considered.

\begin{figure}[htb!]
  \begin{center}
    \includegraphics[width=\textwidth]{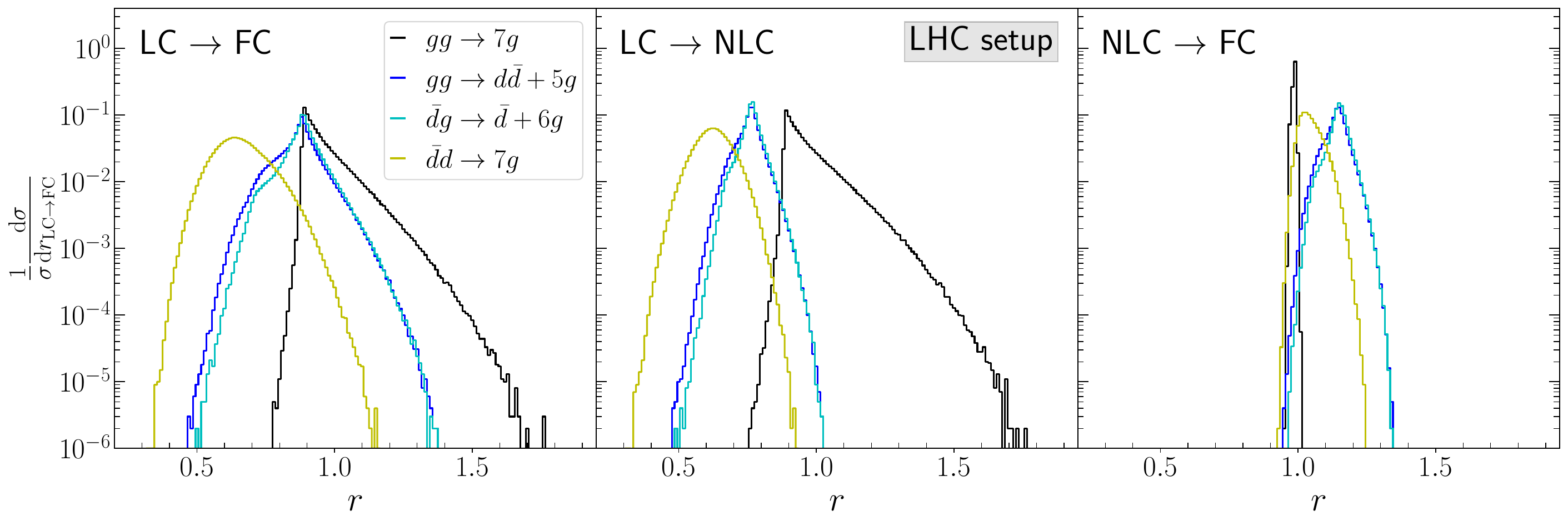}
    \caption{Weight distributions for the reweight factors $r^{\LC\to\FC}$ and
      $r^{\LC\to\NLC}$ and $r^{\NLC\to\FC}$ for the $gg\to 7g$ (black),
      $gg\to d\overline{d}+5g$ (blue), $g\overline{d}\to \overline{d}+6g$
      (cyan), $\overline{d}d\to 7g$ (yellow) processes, for the LHC setup. For each, one
      (representative) colour order is chosen.}
    \label{fig:wgt_dist_1qq}
  \end{center}
\end{figure}

In Fig.~\ref{fig:wgt_dist_1qq} we compare the weight distributions for the
reweight factors for the $n=7$ all-gluon process (in black), with the
one-quark-pair processes: in blue the $gg\to d\overline{d}+5g$ process, in
cyan the $g\overline{d}\to \overline{d}+6g$ process, and in yellow the
$\overline{d}d\to 7g$ process. As expected, for the one-quark-line processes
in which the quarks are not back-to-back, the peak in the $\LC\to\FC$ reweighting
is also located at $(\NC^2-1)/\NC^2$, with a narrower weight
distribution than the all-gluon processes. On the other hand, for the $\overline{d}d\to 7g$ process, the
weight distribution is rather different. The peak is at a lower value and is
much broader, especially close to the peak. We find that this is generally the case when the
kinematics associated with the quark and anti-quark are back-to-back; a
similar broader spectrum is obtained for the $n=4$
$gg\to d\overline{d}$ process (not shown), in which the quarks are also
back-to-back. We have not investigated the mechanism behind this. Comparing
the $\LC\to\FC$ to the $\LC\to\NLC$ weight distributions, we find that the
$\LC\to\NLC$ are considerably narrower, and not peaked at the same values,
which is compensated for by the tertiary $\NLC\to\FC$ reweight
distributions.  Consistent with these weight distributions, in the
$2^{\textrm{nd}}$ to $4^{\textrm{th}}$ columns of
Fig.~\ref{fig:2nd_unw_eff_all_procs} we observe that the secondary
unweighting efficiencies are larger for the $\LC\to\FC$ and $\LC\to\NLC$
reweighting for the one-quark line processes than for the all-gluon process.
On the other hand, the NLC approximation more closely matches the FC result for the
all-gluon process, resulting in a narrow peak at 1 for the $\NLC\to\FC$
reweight factors; for the one-quark-line processes the $\NLC\to\FC$ weight
distribution is wider, but still quite a bit narrower than for the $\LC\to\FC$
(and $\LC\to\NLC$), which confirms the larger tertiary than secondary
unweighting efficiencies for these processes. For lower multiplicities (not
shown)
the weight distributions become considerably narrower, resulting in
considerably larger
secondary and tertiary unweighting efficiencies.

\begin{figure}[htb!]
  \begin{center}
    \includegraphics[width=\textwidth]{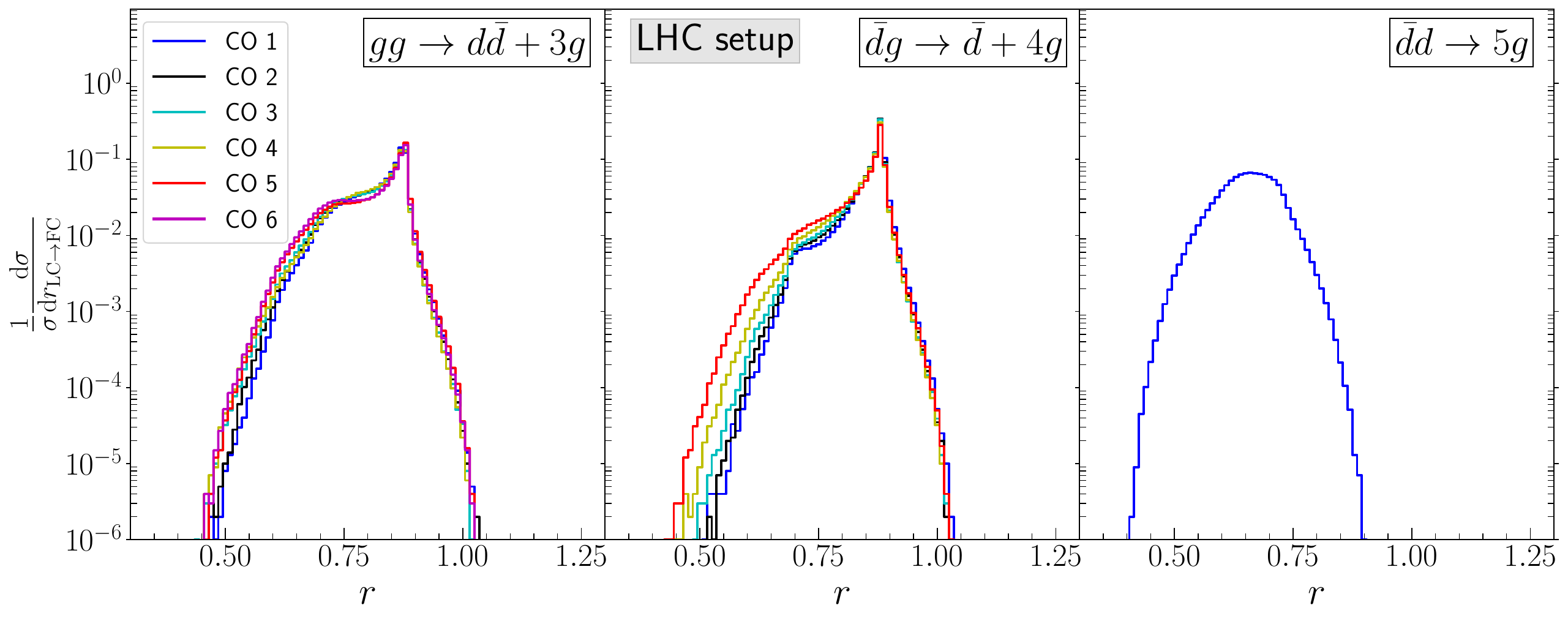}
    \caption{Weight distributions showing the colour ordering dependence for
      the $\LC\to\FC$ reweight factors for the three one-quark-line processes, for the LHC setup.}
    \label{fig:wgt_dist_1qq_co}
  \end{center}
\end{figure}

Before moving on to the two-quark-pair processes, in
Fig.~\ref{fig:wgt_dist_1qq_co} we show the weight distribution for the $\LC\to\FC$
reweight factors for the $n=5$ one-quark-line processes for each of the
distinct colour orderings. Since the only way the colour ordering enters the
weight distributions is through the kinematics of the unweighted LC events, it
can be expected that the dependence of the distributions on the ordering is mild. Indeed, for both
the $gg\to d\overline{d}+3g$ and $g\overline{d}\to \overline{d}+4g$ the
difference is small, whereas for the
$\overline{d}d\to 5g$ process, only one distinct colour ordering needs to be
considered because all final-state particles are identical
gluons (see Tab.~\ref{tab:proc}). We observe the same independence from the
colour ordering also for other multiplicities and for the all-gluon processes;
c.f.~Fig.~\ref{fig:2nd_unw_eff_all_procs}.

In general, for the two-quark-line processes the secondary unweighting
efficiencies in the reweighting of the $\LC$ events to $\FC$, we find unweighting
efficiencies in the 50-70\% range for the $n=7$ processes (lower
multiplicities yield larger efficiencies), see
Fig.~\ref{fig:2nd_unw_eff_all_procs}. Just as in the one-quark-line case,
the unweighting efficiencies for reweighting $\LC\to\NLC$ are about 7-10
percent points higher than those for reweighting to FC. However, there is a significant variation in the
secondary unweighting efficiencies among the different flavour
assignments. Moreover, contrary to the all-gluon and the one-quark line
processes, there is also a dependence in the efficiencies from the colour
ordering in some of the processes. In the latter case, instead of showing a
single representative colour ordering, we present two of them in
Fig.~\ref{fig:2nd_unw_eff_all_procs}. As discussed before, the colour
ordering does not directly enter the reweighting, since both the numerator and
denominator in Eq.~\eqref{eq.rwfactor} are summed over all colour
orderings. It enters only through the kinematics of the events, which, in
turn, is dominated by how the two quark pairs are ordered:
\begin{enumerate}
\item For processes that have a quark-anti-quark pair in the initial and in
  the final state, the ordering in which initial-state anti-quark is followed
  by the initial-state quark and the final-state quark is followed by the
  final-state anti-quark, e.g.~$(b \ldots a3\ldots 4)$ for the process $d \overline{d} \rightarrow u \overline{u} + (n-2) g$, has a larger secondary unweighting efficiency than the
  colour orderings where the initial-state anti-quark is followed by a
  final-state anti-quark and the final-state quark is followed by the initial-state
quark, e.g.~$(3\ldots ab\ldots 4)$.
\item For different-flavour quark line processes, and as discussed in
  Sec.~\ref{sec.LC}, we include all the elements on the diagonal
  of the colour matrix (in the fundamental basis) as part of our LC approximation,
  even though only half of them contribute at LC and the other half only at
  NLC accuracy. The channels for which the colour ordering contributes only at NLC
  (i.e., the quark anti-quark pairs of the same flavours form a substring in the
  colour ordering---possibly with gluons in between) have a smaller secondary
  unweighting efficiency than channels that contribute at LC (i.e., the quark
  anti-quark pairs that form a substring in the colour ordering have different
  flavours).
\end{enumerate}
These two general rules explain the differences observed in the secondary
unweighting efficiencies for the two-quark-line processes: for processes
$g g \rightarrow d \overline{d} d \overline{d} + (n-4) g$,
$d g \rightarrow d d \overline{d} + (n-3) g$ and
$d d \rightarrow d d + (n-2) g$ we find hardly any variation among the
contributing colour orderings and we show the results for only a single
representative colour ordering; for the
$d \overline{d} \rightarrow d \overline{d} + (n-2) g$ process we have two
representative sets of efficiencies due to point 1 above; for processes
$g g \rightarrow d \overline{d} u \overline{u} + (n-4) g$,
$d g \rightarrow d u \overline{u} + (n-3) g$ and
$d u \rightarrow d u + (n-2) g$ we have two sets of efficiencies due to point
2 above; and, finally, points 1 and 2 both apply to the
$d \overline{d} \rightarrow u \overline{u} + (n-2) g$ processes. In the latter
case, the points counteract each other, resulting in a single representative
secondary unweighting efficiency (see the two similar efficiencies for this process in Fig.~\ref{fig:2nd_unw_eff_all_procs}).

\begin{figure}[htb!]
  \begin{center}
    \includegraphics[width=\textwidth]{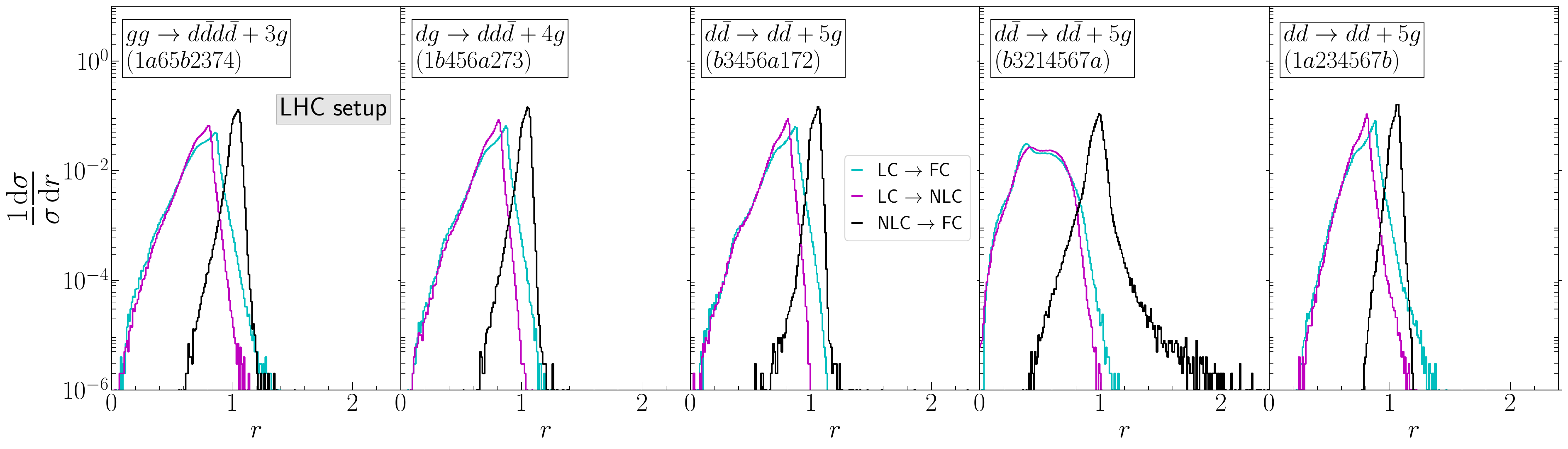}
    \includegraphics[width=\textwidth]{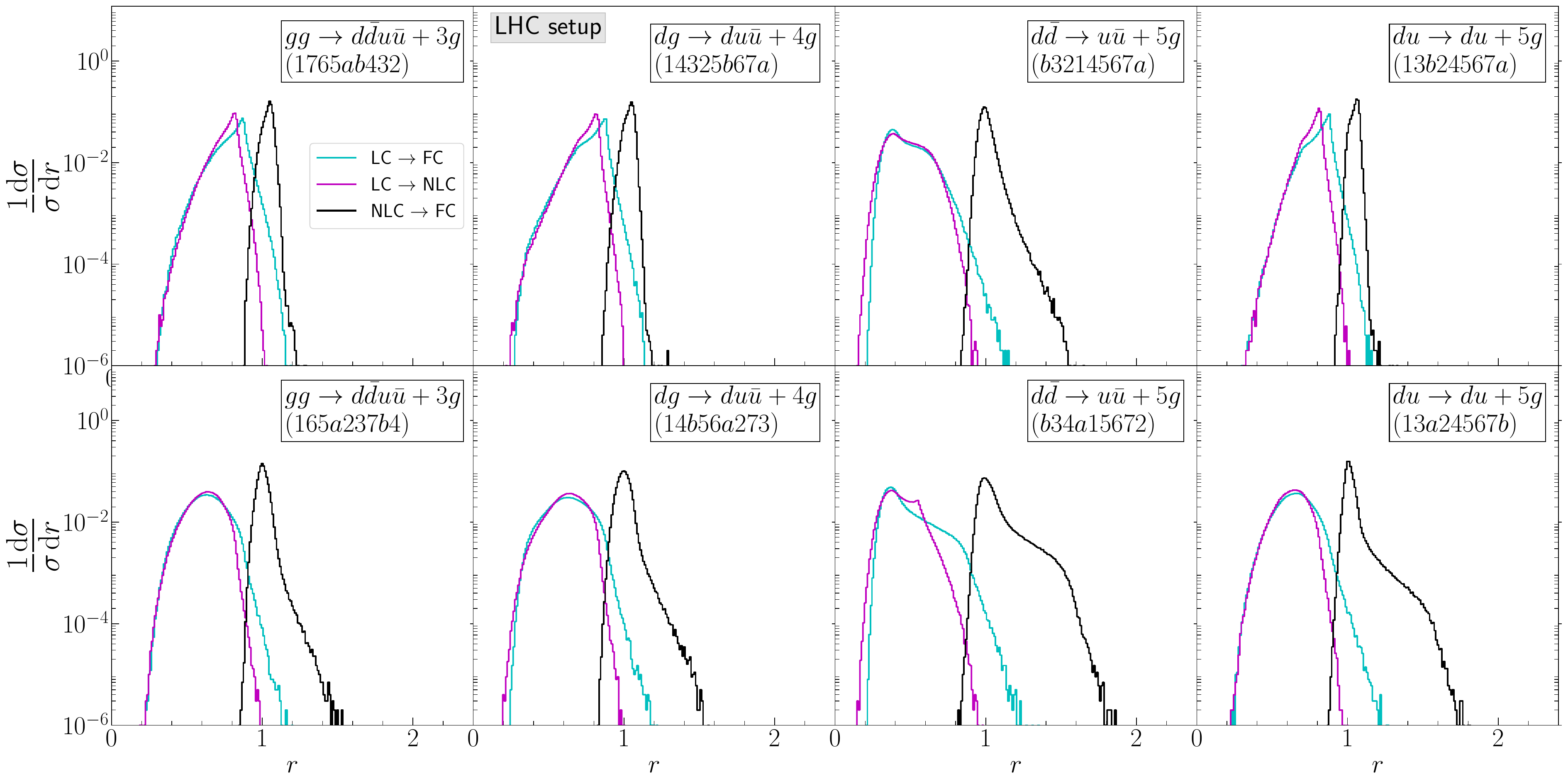}
    \caption{Weight distributions of the reweight factors for
      representative channels in the $n=7$ two-quark-line processes, for same-flavour processes (upper plot) and different-flavour processes (lower plot).}
    \label{fig:wgt_dist_2qq}
  \end{center}
\end{figure}

The differences in the secondary unweighting efficiencies can, of course, also
be observed in the weight distributions for the reweight factors directly, see
Fig.~\ref{fig:wgt_dist_2qq}. In this figure the distributions correspond to the colour orderings for which the efficiencies are presented in Fig.~\ref{fig:2nd_unw_eff_all_procs}. In the top row of this figure, the weight
distributions for a representative channel for the four $n=7$ same-flavour
two-quark-line processes are shown; for the
$d \overline{d} \rightarrow d \overline{d} + 5 g$ process the two
representative quark
orderings, see point 1 above, are presented. In the lower plot, the
different-flavour two-quark-line weight distributions for the reweighting are
shown for both ways the quarks can be ordered, as discussed in point
2 above, with the upper row containing the ordering such that it contributes to LC, and
the lower row such that it contributes to NLC. In cyan we show the weight
distributions of the reweight factors in the $\LC\to\FC$ reweighting, in
magenta for the $\LC\to\NLC$ reweighting and in black the $\NLC\to\FC$
reweighting.

Let us focus on the $\LC\to\FC$ and $\LC\to\NLC$ reweighting first.
As is clear from these plots, the weight distributions in subplots 1, 2, 3, and 5
in the upper plot, and 1, 2, and 4 in the upper row of the lower plot show
nicely peaked distributions resulting in the larger secondary unweighting
efficiencies. The 4th subplot in the upper plot and the 3rd subplot in the upper
row of the lower plot show similar shaped distributions, confirming that the
reduced secondary unweighting efficiency is indeed due to the ordering of the initial-
and final-state (anti-)quarks. The main difference in the reweight factors for
the same-flavour and different-flavour processes is that for the
same-flavour processes, the distribution extends to much smaller values. 
Subplots 1, 2 and 4 in the lower row of the lower
plot again show similar shape; these are obtained from events that were
generated with a colour ordering that only contributes starting at NLC. The
third subplot in the lower row of the lower figure has its own unique shape. As
explained before, this flavour configuration is unique because it is obtained
from an event file generated from a NLC colour ordering, but also has the
initial-initial plus final-final (anti-)quark ordering. 

In the three-step approach, reweighting $\LC\to\NLC\to\FC$, we can see in
Fig.~\ref{fig:wgt_dist_2qq} that the total rate of the $\NLC$ requires a
shift to get to $\FC$ in some of the channels: when the $\LC\to\FC$ and
$\LC\to\NLC$ distributions are very peaked, their peaks do not align, resulting in a
$\NLC\to\FC$ peak that is not located at 1. On the other hand when the weight
distributions are broader, the $\NLC\to\FC$ peak is located closer to 1.

For some of the same-flavour processes (upper plot in
Fig.~\ref{fig:wgt_dist_2qq}), both the $\LC\to\FC$ and $\LC\to\NLC$ reweight
factors approach zero.
Even though the absolute difference between these
two reweight factors is very small in this region, the relative difference can be
significant. Indeed, in the ratio of these factors, i.e., the
$\NLC\to\FC$ reweight factors, we observe a sizable tail towards larger
factors. These runaway tails hamper the tertiary unweighting efficiency
significantly.  Given that we find several configurations with significantly
lower tertiary reweighting efficiencies, and, in particular observe runaway tails
at times, the reweighting (and unweighting) to an
intermediate $\NLC$ accuracy does not appear to be most-efficient for these processes.

\subsubsection*{General remarks}
We end this section with a few remarks regarding the accuracy of first
generating LC events, reweighting them to higher accuracy, and performing a
secondary unweighting.

The first remark relates to the computation of $r_{\textrm{max}}$ used as
maximum value against which the secondary unweighting is performed, see
Eq.~\ref{eq:2ndunw}. Currently we estimate this channel-by-channel by taking
the maximum value found in each event file generated. An alternative would be
to take the maximum value in any of the events generated for a given process,
irrespective of the colour ordering used to generate the individual event
files. In general, the latter method would reduce the unweighting efficiencies
somewhat, but would not change the overall results presented here. On top of
that, we believe that defining $r_{\textrm{max}}$ channel-by-channel, as we
do, is more correct: even though each channel covers the full phase space, due to
the different phase-space parametrisation and integrand, each channel covers
the phase space differently. Phase-space regions for which $r_{\textrm{max}}$
is large can have a much smaller measure in some channels than others,
effectively reducing the $r_{\textrm{max}}$ to use in practice. By keeping a
separate $r_{\textrm{max}}$ in each channel, we take this effect into
account. 

Second, for our procedure to work, the LC approximation must fill the full
phase space: it must not be the case that for a given phase-space region the
LC matrix elements are zero (or very small) while the FC ones are
non-zero. For the processes considered here there is indeed no sign of such
phase-space regions: there are no runaway tails in the distributions of the
reweight factors that hamper the secondary unweighting efficiencies: they all
have a relatively sharp drop-off above which there are no events. This,
together with the fact that the cross sections agree with known results within
statistical uncertainties shows that we have obtained the correct FC results
with our method.

On the other hand, for the $\NLC\to\FC$ reweighting for the same-flavour
two-quark-line processes, these phase-space regions do exist. For some
channels we indeed notice that there are phase-space regions in which the
$\NLC$ is (close to) zero, while the $\FC$ is not, resulting in a large
variation in the tertiary unweighting efficiencies among the 10 different random
seeds used for each channel. Moreover, one can also directly observe this in
the weight distributions for the $\NLC\to\FC$ reweight factors for these processes. This means
that one should not perform a three-step approach, but rather reweight LC
directly to FC, where this problem does not arise. Alternatively, it might be
possible to include some terms beyond the NLC accuracy in our NLC
approximation to move it closer to the FC and by doing so removing the
phase-space region where the NLC approximation is close to zero.

\section{Conclusions and Outlook}\label{sec:conclusions}
In this work we have presented a two-step approach to event generation at
leading order in the perturbative expansion. In a first step we use a fast approximation of the integrand to
perform the phase-space integration and generate unweighted events. We use (a
small number of) single colour-ordered matrix elements as the approximation;
each single colour-ordered amplitude squared forms its own channel and can
be integrated in parallel. Since the integrands are very fast to evaluate, and
also have a relatively simple phase-space dependence, the generation of these
events is fast and the unweighting can be performed at high efficiency: even
for $2\to 7$ processes, we find efficiencies at or above the per-mille
level with the \gen\ phase-space parametrisation in all the 
integration channels.

As a second step, the LC-accurate events are reweighted to full-colour,
and unweighted again. Since phase-space
dependence in the needed reweight factors is small, secondary unweighting
efficiencies are typically well above 50\%, even at the highest multiplicities
we have considered in this work. In practice this means that in order to generate a
certain number of unweighted events, for less than twice that number the slow,
full-colour matrix elements need to be evaluated. Moreover, this reweighting
procedure can be trivially parallelised, since each LC event can be reweighted
independently.

Compared to the traditional approach of integrating the full-colour matrix
elements directly over the phase space (either with or without a Monte Carlo
sum over colours), our method yields a double improvement. Not only is the
number of phase-space points for which the full-colour matrix elements need to
be evaluated much smaller in the generation of a given number of unweighted
events, also the phase-space integration itself is more efficient since the LC
matrix elements for a given colour ordering have a much simpler dependence on
the phase space than the full-colour ones.

We plan to apply our method also to processes including massive QCD particles
such as top quarks or bottom quarks and electroweak bosons in the future. We
do not foresee any problems for the inclusion of masses; the inclusion of
non-coloured particles might impose more difficulties. In particular, our
phase-space generation for the LC events is optimised for a specific colour
ordering of all the produced particles---non-coloured particles do not
fit neatly into this picture, and might require several more integration channels for
efficient event generation and unweighting. On the other hand, electroweak particles do
not play any special role in the reweighting of the LC events to FC, and so
the second step in our approach should not be seriously affected.

It would be interesting to investigate if the method of using several
consecutive unweighting steps can be applied more broadly. For example,
starting from an unweighted LO event file, could the events be efficiently
reweighted to generate (part of) an NLO-accurate event file? In particular,
the $\mathbb{S}$-events in MC@NLO~\cite{Frixione:2002ik} or
MC@NLO-$\Delta$~\cite{Frederix:2020trv} matching, or the $\overline{B}$
function in the POWHEG method~\cite{Nason:2004rx}, have the same kinematics as
the LO contributions; they could be suitable candidates for studies in this
direction.

\section*{Acknowledgements}

This work was supported by the Swedish Research Council under
contract numbers 201605996 and 202004423.  The work of T.V. is
supported by the Swedish Research Council, project number
VR:2023-00221.

\bibliography{Paper}{}
\bibliographystyle{JHEP} 
\end{document}